\documentclass[article, 10 pt, conference]{ieeeconf}  

\IEEEoverridecommandlockouts                              
\usepackage{graphics} 
\usepackage{epsfig} 
\usepackage{mathptmx} 
\usepackage[export]{adjustbox}
\usepackage[numbers]{natbib}

\title{\LARGE \bf
The never ending war in the stack and the reincarnation of ROP attacks
 }

\author{ Ammari Nader, Joan Calvet, Jose M. Fernandez}

\begin{document}

\maketitle
\thispagestyle{empty}
\pagestyle{empty}

\begin{abstract}

Return Oriented Programming (ROP) is a technique by which an attacker can induce arbitrary behavior inside a vulnerable program without injecting a malicious code. The continues failure of the currently deployed defenses against ROP has made it again one of the most powerful memory corruption attacks. ROP is also considered as one of the most flexible attacks, its level of flexibility, unlike other code reuse attacks,  can reach the Turing completeness. Several efforts have been undertaken to study this threat and to propose better defense mechanisms (mitigation or prevention), yet the majority of them are not deeply reviewed nor officially implemented. Furthermore, similar studies show that the techniques
proposed to prevent ROP-based exploits usually yield a high false-negative rate and a higher false-positive rate, not to mention the overhead that they introduce into the protected program. The first part of this research work aims at providing an in-depth analysis of the currently available anti-ROP solutions (deployed and proposed), focusing on inspecting their defense logic and summarizing their weaknesses and problems. The second part of this work aims at introducing our proposed Indicators Of Compromise
(IOCs) that could be used to improve the detection rate of ROP attacks. The three suggested indicators could detect these attacks at run-time by checking the presence of some artifacts during the execution of the targeted program. We also proposed a measurement technique that allows measuring these indicators at run-time.
The last part of this work covers the subject of the experimental phase. More specifically, the Proof of Concept performed with the objective of proving the effectiveness of our proposed indicators, as well as the proposed measurement technique. The results of this experimental phase show that only the first two indicators are able to detect ROP attacks. Another important finding was about the non-expected ROP features discovered and visualized during the experiment. These features could be used to strengthen our indicators in future works.
\end{abstract}

\section{INTRODUCTION}

Over the last few years, computer programs have become an integral part of our society. Unfortunately, these programs are known to contain some errors and imperfections which could be exploited by attackers to achieve malicious operations. The most worrisome scenarios are those during which an error is exploited with the goal of gaining full control over the program. In this case, the attacker becomes able to fully control the targeted program, and may as well control the machine that executes it and all the connected devices. From a technical perspective, these attacks usually rely on corrupting the program's memory to force the execution to deviate and execute the malicious code instead of the program's code\cite{memcorruption}. Traditionally, attackers have been achieving this by injecting the malicious code somewhere in the program's memory before deviating the execution towards it. These attacks are referred to as ``code injection attacks''. The injection method varies from one attack to another depending on the situation and the attacked program. Well known examples of this include injecting the malicious code inside the user inputs. More creative injection techniques could even involve the use of  side-channel attacks such as relaying on electromagnetic waves or sound waves to remotely deliver and inject the malicious payload\cite{sidechanel}. A profusion of defense mechanisms has been proposed, engineered and implemented to stop this kind of ``code injection attacks''. To date, the most effective defense against code injection attacks is known as ``Data Execution Prevention'' (DEP)\cite{DEP}. This defense prevents code injection attacks by simply making some memory locations non-executable. By doing this, it can guarantee that if an attacker injects any code, he will never be able to execute it. DEP has been integrated not only by the majority of operating systems but also by hardware manufacturers. The great success of DEP made hackers unable to perform traditional ``code injection attacks''. However, the wide adoption of DEP has forced attackers to invent and use new techniques, one of these newly engineered techniques proposes  \textbf{reusing} code and functions from inside the targeted program to build the malicious payload instead of injecting new code\cite{codereuseattacks}\cite{codereuseattacks2}, hence the name ``code reuse attacks''. The possibility of reusing a program's code to build an attack was both a shocking fact and an unexpected trick that may be compared (strategically) to a ``cancer'' where body cells start attacking themselves. The tremendous success of code reuse attacks made them one of the top security concerns and one of the most potent cross-platform weapons\cite{codereuseattacks}. Originally, these code reuse attacks focused on re-utilizing the code of whole functions. However, from an attacker perspective, reusing the entire function was not providing enough flexibility to build the desired malicious code. For that reason, a new customized version of code reuse attacks was designed. This version dubbed ``Return Oriented Programming attacks''(ROP) \cite{shacham2007geometry} proposes reusing small pieces of code instead of reusing an entire function's code. These small chunks of code could be located inside libraries code (loaded by the program) or the program's code, and they are chained together using the Ret instruction \cite{x86assembly} which should be located at the end of each chunk. To date, ROP is considered as one of the most sophisticated and flexible ``code reuse attacks''\cite{ropdanger}. Furthermore, \citet{turingrop} proved that the level of flexibility offered by ROP attacks could achieve the Turing completeness, making attackers able to perform more powerful attacks and obviating the need for injecting codes. This powerful code breaking strategy inspired hackers to create similar versions like ``Jump Oriented Programming'' (JOP) attacks\cite{jop}.

\section{ The primary defenses are collapsing }

As a solution to mitigate memory corruption attacks, a new defense mechanism was designed and implemented inside the majority of operating systems. This mechanism is known as ``Address Space Layout Randomization'' (ASLR)\cite{aslr}. ASLR was originally designed to mitigate the old versions of code injection, however, it does also make ROP attacks harder to exploit. Based on the research work provided by \citet{ropdanger}, ASLR could be considered as the first defense line against ROP attacks. The main idea behind ASLR is to load the program's components in random memory locations, by doing so, attackers are no longer able to reuse program's code as they cannot predict its new location. Unfortunately, ASLR is no longer able to mitigate memory corruption attacks as attackers have been able to design new circumvention techniques. The most straightforward bypassing technique consists in finding a non-protected library loaded by the targeted program. This problem is manly related to the default settings of the widely used development tools as they do not activate the randomization by default during the compilation process. As an example, in Visual Studio 2008, the developer has to explicitly add the flag ``/DYNAMICBASE'' in order to compile an ASLR-protected module. Furthermore, in Windows 8, the developer has to specify a second flag named ``/HIGHENTROPYVA'' in addition to ``/DYNAMICBASE'' just to ensure that the entropy of the randomization is high, this limitation made brute-force attacks one of the most used techniques to bypass ASLR. While testing the effectiveness of brute-forcing a function's address, the research team learned that the randomization for some operating systems such as ``Windows 8'' is limited to 8 bits. In other words, an attacker can guess the protected address after trying at maximum $2^8 = 256$ tries. Memory leak attacks are also used to bypass ASLR. From a general perspective, hackers could rely on a format string attack or simply forcing the program to print an error message to extract/reveal the protected information. Return-to-plt attack is another bypassing technique used to circumvent ASLR. During this attack, the attacker calls the needed function using its PLT\cite{ret-to-plt} instead of its address, by doing so, he is no longer forced to guess the unknown addresses. More advanced ASLR circumvention techniques propose relying on hardware weaknesses which, unlike software weaknesses, are known to be hard if not impossible to patch. A well-known example of these hardware-based circumvention techniques is the ``ASLR and Cache'' (AnC)\cite{AnC2017} attack, introduced to the public in late 2016. As its name indicates, AnC relies on the cache hierarchy of the processor to disclose the needed addresses and bypass ASLR. The effectiveness of AnC attack has been proved against 22 CPU microarchitectures\cite{AnC2017}, including the most recent architectures designed by Intel, AMD, Samsung, Nvidia, and Allwinner. In addition, another powerful hardware-based circumvention technique was proposed in late 2016 by \citet{aslr-bypass2}. This technique exploits a weakness in the branch target buffer (BTB) which is used by modern processors to predict the target of a branching operation. In their research paper, \citeauthor{aslr-bypass2} explained how an attacker could bypass a Kernel ASLR protection in 60 milliseconds by exploiting this weakness in BTB. The apparition of the previously described attacks (i.e. AnC and BTB attack) represented the end of ASLR protection, leaving the cyberspace without any effective defenses against advanced memory corruption attacks such as ROP. Since then, the number of the successfully exploited ROP attacks is becoming higher every year as shown in the reports provided by the commune vulnerability and exposure database\cite{cve}. The failure of ASLR defense is not the unique worrisome fact, the second frightening fact is about how attackers are escalating the use of ROP attacks to become a cyber-espionage weapon used in advanced targeted attacks and state-sponsored attacks. A well-known example of this is the Operation Clandestine Wolf\cite{clandestine-wolf} performed by a China-based threat group known as ``APT3''. During this operation, a ROP attack was used to launch a large-scale spying campaign against aerospace companies, defense organizations, construction companies, engineering companies, high-tech malefactors, and even transportation companies. Operation Greedywonk\cite{greedywonk} is another example of rop-based cyber-espionage campaign, during which, hundreds of economic and foreign policy sites were compromised for more than two years, and were spreading the infection to some targeted high-profile personalities without being detected. In the absence of effective defenses against ROP attacks, and because of the previously-stated escalation in its use, ROP attacks may be viewed as a serious security threat. The existence of such threats heightened the need for new defenses. Therefore, academia has shown an increased interest in studying ROP attacks and proposing new alternatives and solutions. A quick scan of the literature confirms that a profusion of research on ROP attacks has been published, though the majority of them have not been deeply reviewed or officially implemented. 

\section{The secondary defenses are not holding well }

\subsection{STACK COOKIE}

Stack Cookies\cite{stackcookies} is one of the oldest defense techniques against memory corruption attacks. It is nearly supported by every operating system (e.g. Windows, Linux and Mac Os ) and usually referred to as “canaries protection”. This protection is implemented at the compiler level, and it is focusing on securing the return addresses. During the compilation process, the compiler adds some extra pieces of code to the protected function. The main task of the injected codes is to guarantee the integrity of the stack. More specifically, the protection is performed by injecting a randomly generated 4 bytes known as a “cookie”, this cookie is usually injected between the local variables and the return address of the protected function. The original copy of this cookie is called program-wide master cookie which is saved in the “.data” segment. During the function’s epilogue, the injected cookie is compared again with the original one. If the two compared values are not matching the operating system concludes that the integrity of the stack is violated. From an attacker’s perspective, it is impossible to overwrite the return address without corrupting the value of the injected cookie. The main reasons behind the failure of this security mechanism are the deployment cost and the restrictions applied by the majorities of the compilers. Previous studies have confirmed these facts, such as the one conducted by \citet{stackshadow-and-cannariesp-perform}. \citeauthor{stackshadow-and-cannariesp-perform} explained in their research paper that the deployment of stack cookies comes with a high price, additional codes are being injected into every protected function, forcing the program to execute additional calculations at the beginning and the end of every function, without mentioning the additional stack storage reserved to save all cookie values. Furthermore, the restrictions imposed by compilers are playing an important role in this failure, in other words, the majority of the compilers are not activating this protection in so many cases such as when the protected function is not declaring any stack buffers, when a function is marked with “naked”, when the argument list of the function is variable or when the optimization is disabled. From an attacker's perspective, the easiest way to bypass Stack Cookies protection is to start by auditing the targeted program in order to locate some of these non-protected functions. Even if the program is well protected and all its internal functions are protected, hackers still able to exploit these exceptions by relying on external vulnerable dependencies (e.g. loaded libraries) in order to find these non-protected functions. Another bypassing technique proposes the use of “brute-force attack” and trying all the possible combinations. On a 32bit system, the canary field is composed by four random bytes, however, the first one is always null \cite{canary-brute-force}, making the maximum number of trials equal to 8 388 608. Based on the study conducted by \citet{canary-brute-force}, this number of combinations could be brute-forced in few hours. Furthermore, it has been proven that stack cookies could be brute-forced in less than one minute using an advanced brute-force attack known as ``Byte-for-byte brute force''\cite{byte-for-byte-stack-cooki}. During this attack, the number of the needed trials is decreased to 768, \citet{canary-brute-force} discussed this attack in their research paper and said: "The attack consists in overwriting only the first byte of the canary until the child does not crash. All the values from 0 to 255 are tested sequentially until a success is got. The last byte tested is the first byte of the canary. The remaining bytes of the canary are obtained following the same strategy. This kind of bug is very dangerous because a system is broken with only 3\*256 = 768 trials. Finally, memory leak attacks could also be used to circumvent Stack Cookies protection. As an example, the attacker can use a ``format string attack''\cite{format-string} to simply extract the value of the cookie from the stack.

\subsection{ASCII ARMOR}

During code reuse attacks the attacker has to specify the destination address after hijacking the control flow of the program, this is usually done by injecting the destination address after the payload that will trigger the overflow and overwrite the instruction pointer value. The destination address can not contain any null-byte character (i.e. “00”) as the majority of the vulnerable functions are assuming that their inputs should terminate with a null-byte charter. ASCII-Armor is exploiting the previously described fact to mitigate ROP attacks and other similar code reuse attacks. When this mitigation technique is enabled, all the system libraries (e.g. Libc) are loaded in a particular addresses that contains null-bytes, making the attacker unable to reuse functions or gadgets from these libraries. The main problem of ASCII-Amor protection, is that it can be easily circumvented using PLT tables \cite{ret-to-plt} and use a function's PLT instead of its address\cite{ret-to-plt}. Another bypassing technique consists in using machine instructions to build/construct the protected address without explicitly using the null character. Further more, it is always possible to find several non protected libraries.

\subsection{DATA EXECUION PREVENTION }

The first approach adopted by attackers to defeat Data Execution Prevention (DEP) is to exploit some specific functions able to modify the DEP policy. On Windows platforms, the best known DEP bypass function is VirtualAlloc. \citet{dep-virtual-alloc} demonstrated the use of VirtualAlloc to bypass DEP, the attack consists on using this function to create a new executable memory region where the malicious code could be injected and executed. In other words, it allows the creation of non-protected memory space. HeapCreate is also a similar function that provides the same privileges as VirtualAlloc. Furthermore, there are so many similar functions such as VirtualProtect, WriteProcessMemory, SetProcessDEPPolicy, SetProcessDEPPolicy and ZwProtectVirtualMemory, they are also able to manipulate the Data Execution Prevention policy and could be used to bypass this security mechanism.

The second approach used by hackers to bypass DEP proposes the exploitation of the DEP misconfiguration. More specifically, DEP could be functional under four possible modes. The first mode is referred to as OptIn Mode where only a limited set of binaries are protected.
The second mode is referred to as OptOut Mode where all binaries on the system are protected, except the programs specified in the exception list. The third mode is referred to as AlwaysOff Mode where DEP is disabled for all binaries. The forth mode is known as AlwaysOn Mode Where all binaries on the system are protected with no exception. The OptIn Mode is usually circumvented by relying on a non-protected binary. The OptOut Mode is usually bypassed by first gaining access to the exception list, then using one of the binaries specified in this list to perform the attack. When the AlwaysOff Mode is activated, the attacker can use any binary in the system to perform the attack as all of them are not protected. If the AlwaysOn Mode is activated, the attacker can rely on the previously described functions such as VirtualAlloc and Heapcreate to create a non-protected memory space. In addition, all kinds of code-reuse-attacks such as Return Oriented Programming, Jump Oriented Programming, Ret-to-libc or Ret-to-PLT can circumvent DEP as they are not injecting any codes. Actually, code reuse attacks have been engineered with the goal of bypassing this security mechanism.

\subsection{SUMMARIZING THE WAR IN THE STACK}
Figure \ref{fig:the-war-in-the-stac} provides a detailed summary of the  bypassing techniques discussed previously in this work and used by hackers to circumvent the currently deployed security mechanisms, the figure describes the link between all these attacks and how they are chained together to achieve a successful code execution attack.

\begin{centering}

\begin{figure*}[h]
\includegraphics[scale=0.4]{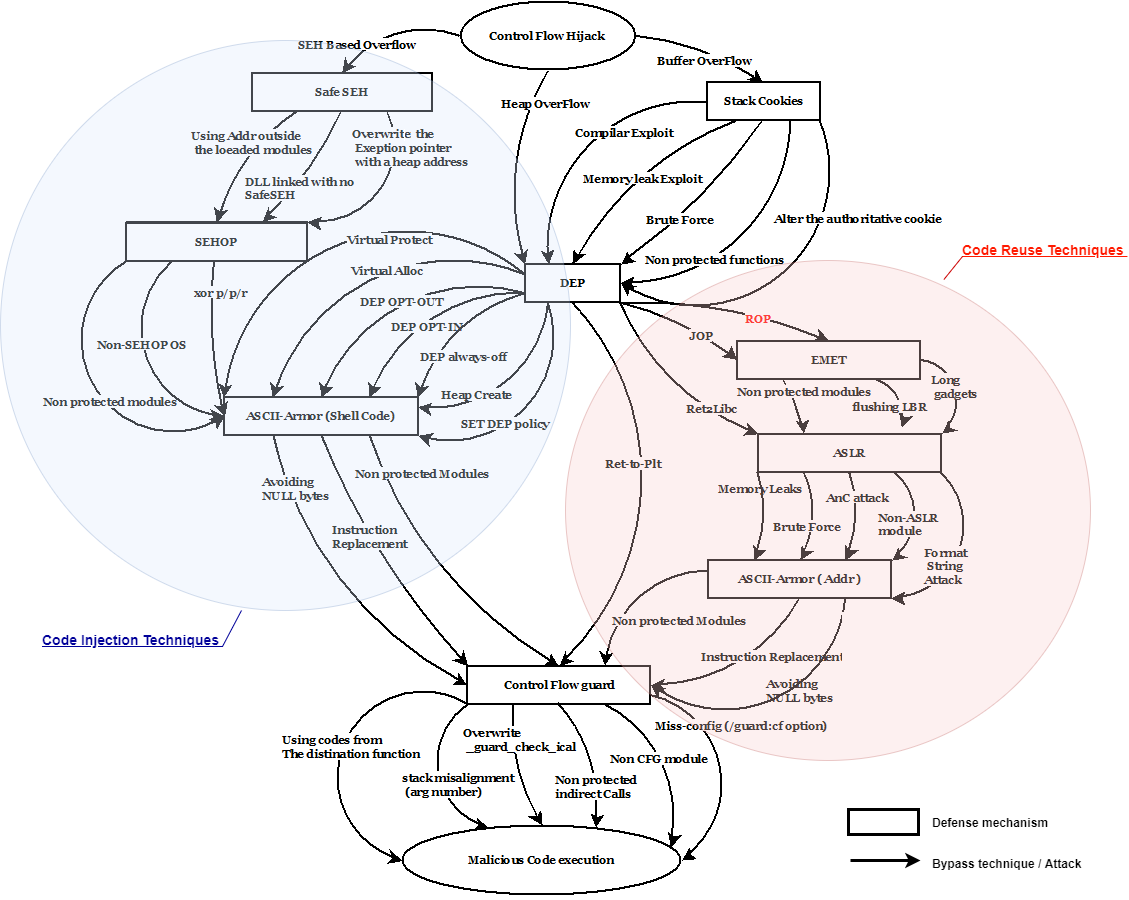}
\caption{The War In The Stack }
\label{fig:the-war-in-the-stack}
\end{figure*}
\end{centering}

\section{ The academic proposed solutions and their limitation }

This section provides a detailed explanation of the most important academic anti-ROP solutions studied during this work, our study aimed at inspecting all the details of these proposed defenses, the inspection went trough all the solutions layers from the logic that they propose to the way how they implemented this logic. The studied solutions were classified in such a way that it becomes easier to compare them, four categories have been proposed based on the approach adopted by the authors. The first category discusses the compiler level approaches where authors are exploiting the compiler capabilities in order to defeat Return Oriented Programming attacks. In this category, two solutions were analyzed and detailed. The first solution was ``G-Free'', which proposes the application of some compiler-based patches in order to prevent the non-authorized branching executed during a ROP attack. The second solution was ``Return-less kernel'', this solution proposes the replacement of the x86 instruction ``Ret'' in order to prevent hackers from reusing them to build the malicious gadgets. The second category lists the solutions where ``dynamic binary instrumentation'' is used to analyze the targeted binaries at run-time. The first analyzed solution was ``ROP-Defender'', it proposes the use of a new concept named ``Stack shadow'', which is a copy of the regular stack used to detect the violation of the stack integrity during a ROP attack. The second solution is named DROP. It proposes translating the execution into an intermediate representation, then inspecting this representation and looking for ROP signature, the proposed signature is similar to the one proposed by our second indicator. However, the manner and logic of counting the instructions are very different. The third category is grouping the solutions where the problem of ROP attacks is treated as a control-flow violation problem. In this category, three solutions have been studied and analyzed. The first solution was ``ROPSTOP'', its main idea consists in making sure that, after executing a subroutine, the control gets back always to the caller. The second solution we studied was ``KBouncer'', which relies on hardware features not only to verify that the control flow is getting back to the right caller but also to the exact right address. The last solution analyzed in this category was ``Control Flow Locking'', as its name indicates, this solution proposes the injection of some locking/unlocking codes inside the protected binary in order to prevent ROP attacks from hijacking the control flow.  The fourth category is about the static binary rewriting approaches where the protected binaries are rewritten in such a way that ROP exploitation become harder. The studied example was ``ILR Instruction Location Randomization'', this solution proposes changing the positions of all the instructions inside the Code segment in order to thwart the hacker's ability to locate the needed gadgets. 

\subsection{OVERVIEW OF THE MAIN LIMITATIONS DISCOVERED IN THE PROPOSED SOLUTIONS}

The inspection of the proposed academic solution revealed several misconceptions problems which make them either exposed to an easy circumvention or hard to deploy. As an example, the encryption used by G-Free\cite{g-free} is too weak and could be broken easily (xor based encryption). Another limitation of G-free is that it violates the integrity of the protected binary which makes its deployment impossible in the case of the digitally signed programs. One of the main limitations discovered in ``Return less Kernel''\cite{retless} was related to the fact that it could be circumvented by simply corrupting the ``Return address table'', in addition, it is impossible to implement such protection in the case of non-open sourced operating systems as it requires recompiling the full kernel. The limitations of ROP-Defender\cite{ropdefender} were related to the adopted assumptions. These assumptions are not correct in the cases of exceptions, signals, interruptions and all goto-like statements which makes its false positive rate very high. A significant limitation was discovered in DROP\cite{drop}, it is related to the use of an abstract form of instructions stream (an Intermediate Representation). Relying on an Intermediate Representation (IR) to inspect the executed instructions will introduce a huge delay and the attack will not be detected in real-time (DROP has to execute the code then translate it to an IR before being able to perform a static inspection of the IR output). The logic adopted by ROPSTOP\cite{ropstop} also suffers from significant limitations, it could be circumvented by using gadgets from the caller subroutine, and the false positive rate is very high. The problems discovered in Ropecker\cite{ropecker} and KBouncer\cite{kbouncer} are mainly related to the use of the Last Branching Records, these records are only able to capture the last 16 branching operations and they could be flushed by running garbage functions as demonstrated by \citet{antirop}. Besides, the two solutions are only protecting very specific functions believed to be ``critical functions'', and the attacker is still cappable of using the non-protected functions to build the ROP chain. Furthermore, KBouncer relies on the DetourDll in order to intercept the execution of the protected functions, many viruses and malware use this exact same DLL to perform malicious actions, therefore it will be flagged as malicious by the majority of the anti-virus products. The solution proposed by CFL\cite{cfl} cannot be applied in the case of packed binaries or self-modifying code as the static extraction of the control flow graph is not always possible in these cases. The solution proposed by ILR\cite{ilr} is violating the integrity of the program and could not be applied in the case of the digitally signed programs.

\section{The proposed indicators to detect ROP}

As explained previously, our analysis of the deployed defenses shows that they are unable to fully detect/mitigate ROP attacks. In addition, the analysis of the academic proposed solutions presented previously shows that these solutions need enhancements as they suffer from many limitations such as deployment problems, high overhead, low detection rate and the majority of them could be easily circumvented. The absence of effective protections against ROP attacks heightened the need for new defenses, thus, we propose three indicators of compromise which could be used to detect ROP attacks at runtime, these indicators could be considered as a meta-data generated unwillingly by the ROP attack, and their existence indicates the presence of such attacks. The first indicator proposes the inspection of every Ret\cite{x86assembly} and Call\cite{x86assembly} instruction being executed by the program. From a general perspective, it relies on the fact that ROP attacks will prevent these instructions from performing their normal tasks\cite{shacham2007geometry}. The proposed indicator is defining this abnormal behaviour and how is could be observed. The second proposed indicator focuses on inspecting every instruction being executed between two consecutive Ret\cite{x86assembly} instructions. The idea is also about detecting the violation of some specific conditions that must hold true when executing codes located between two Ret instructions. The last indicator focuses on the variation of the Instruction Pointer\cite{x86assembly}. More specifically, it proposes the inspection of this variation with the goal of identifying any abnormal patterns that could be introduced by ROP attacks. 

\subsection{Ret/Call parity indicator}

\begin{figure}[h]
\centering
\includegraphics[scale=0.4]{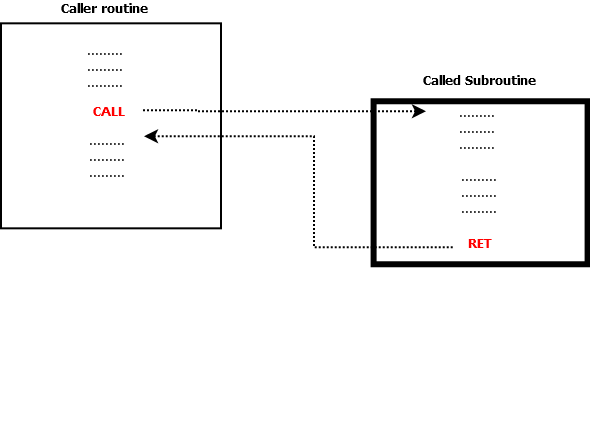}
\caption{ Ret/Call logic }
\label{fig:retcall}
\end{figure}

Programs often rely on subroutines to perform some tasks such as reading users inputs, updating a database or performing calculations. In order to be able to call these subroutines, programs usually rely on the x86 assembly instruction Call\cite{x86assembly}. This instruction interrupts the main execution, saves the address of the interruption onto the stack, then transfers the control to the called subroutine. When a subroutine ends, the program needs to resume the previous execution from where it was interrupted, therefore it relies on the Ret\cite{x86assembly} instruction located at the end of each subroutine to read the previously saved address then jump to it. This mechanism shows perfectly how Call and Ret are always paired together and how “Call” instruction is always executed before the Ret as shown in Figure \ref{fig:retcall}.

\begin{figure}[h]
\centering
\includegraphics[scale=0.6]{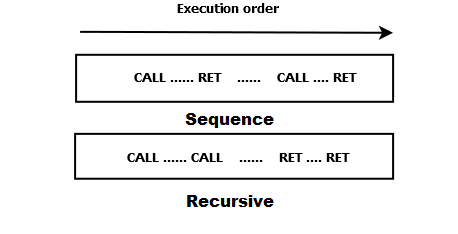}
\caption{ Ret/Call possible execution order }
\label{fig:retcallorder}
\end{figure}

Call and Ret could be executed in two possible ways, the first one is when two subroutines are called one after the other (in sequence). The second way is when the first subroutine calls the second one (recursively). Figure \ref{fig:retcallorder} shows how Call and Ret would be organized in both cases.

\begin{figure}[h]
\centering
\includegraphics[scale=0.28]{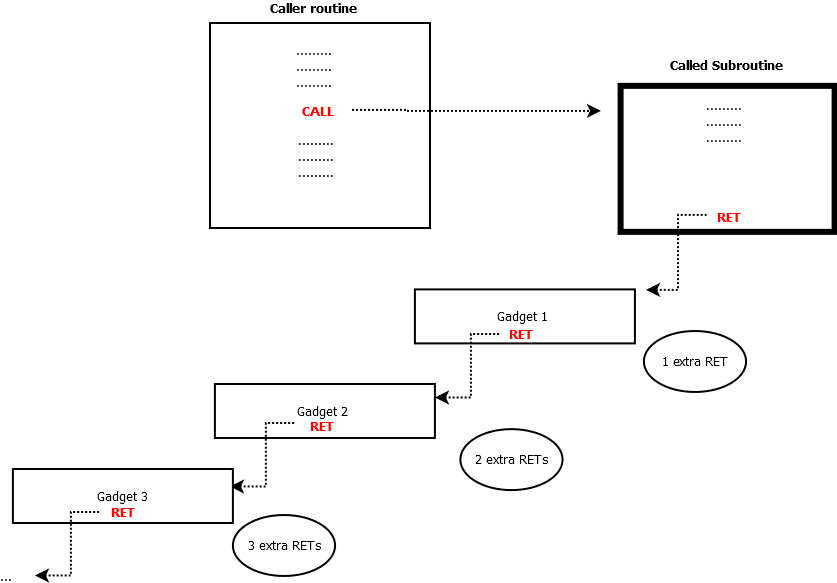}
\caption{ ROP gadgets makes the number of the executed Rets higher than the number of the executed Calls}
\label{fig:ropextraret}
\end{figure}

Based on the previous explanation, it becomes obvious that in both cases and at any moment during the execution, the number of the executed Call should be higher or equal to the number of the executed Ret. However, this fact does not hold true during a ROP attack as it interrupts the normal execution and forces the program to execute gadgets which only contain Ret instructions\cite{shacham2007geometry}. It can therefore be assumed that the number of the executed Ret will increase until becoming higher than the number of the executed “Call” as explained in Figure \ref{fig:ropextraret}.

The proposed indicator exploits the previously described anomaly introduced by ROP attacks to identify the malicious execution. More specifically, we propose monitoring the number of the executed “Ret” and “Call” at run-time and comparing them to make sure that they reflect the legitimate behaviour, where the number of the executed “Call” is always higher or equal to the number of the executed “Ret”.

\subsection{Executed instructions between two Ret }

The number of the instructions executed by a subroutine depends on the complexity of the performed tasks, the manner in which it was implemented and the optimization performed by the compiler during the compilation process\cite{compiler-optim}. However, nearly any subroutine must contain at least five to ten necessary instructions, these instructions are responsible for first preparing the stack when the subroutine starts, second cleaning the stack and restoring it to its original state when the subroutine ends. These processes are known as “function prologue”\cite{x86assembly} and “function epilogue”\cite{x86assembly}.

\begin{figure}[h]
\centering
\includegraphics[scale=0.4]{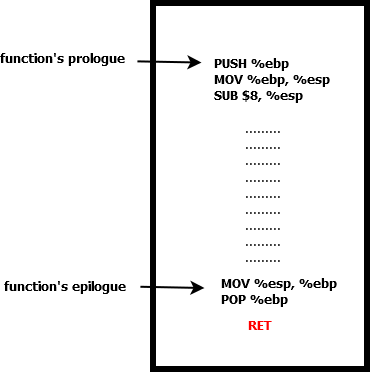}
\caption{  Function Prologue and Epilogue }
\label{fig:epilog}
\end{figure}

Figure \ref{fig:epilog} represents an example of a typical prologue and epilogue, the first three lines of code are the function prologue, it starts by pushing the current base pointer “ebp”\cite{x86assembly} onto the stack so that the program can restore it later. The second line in the prologue writes the value of the stack pointer “esp”\cite{x86assembly} into the “ebp” in order to create a new stack frame on top of the old one. The third line of code is taking care of creating the space for the function's local variables. Depending on the architecture, the “esp” could be decreased or increased, if the stack is growing down, which is the case for the x86 arch, then the “esp” should be decreased as shown in the third line of the prologue. The last two lines of code are the function epilogue, this is where the stack gets cleaned. The first line of code in the epilogue restores the previously saved “esp”, while the second line is taking care of restoring the value of the base pointer “ebp”. In addition, besides the prologue and the epilogue, a subroutine must contain some core instructions, they are related to the execution of the main tasks, and they should be located between the prologue and the epilogue. The total number of the executed instructions in a subroutine is composed of these core instructions plus the epilogue and the prologue. Based on the facts presented thus far, the total number of the executed instructions between two Ret should not be so small and must always stay a bit high, more specifically higher than five. However, this does not hold true during a ROP attack since it is executing a small number of instructions\cite{shacham2007geometry}. According to \cite{shacham2007geometry}, this number could vary from three to five instructions. This indicator proposes checking the number of the executed instructions between two successive Ret and making sure that this number is higher than five, if not, then it shouldn't stay less than five for more than three constitutive Ret as the minimum number of gadgets that can be found in a ROP chain is three\cite{drop}. It is also important to point to the fact that there are some exceptions where a subroutine does not contain an epilogue or a prologue, such as “leaf functions” and “naked functions”\cite{naked}. If the number of the core instructions executed by these functions is so small, they may introduce false positive alerts in this indicator.

\subsection{Instruction pointer variation} 

The code of a program is composed of a set of instructions. At run-time, they are loaded in a specific memory location known as "code segment" (CS)\cite{x86assembly}. More specifically, the ``loader'' reads the contents of the executable and writes it into the code segment from where they will be executed. These instructions are executed one by one, after executing every instruction the program decides what should be the next instruction and jumps to its address, this address is always stored in a register (processor register) called the Instruction Pointer (IP). This pointer is updated after executing any instruction\cite{x86assembly}. It is a widely held view that the variation of this pointer, the distribution of the executed instructions and the order and the time of their execution is completely random, but this is not true. In fact, they obey to some fundamental principles such as the ``Temporal locality'' and the ``Spatial locality''. These two principles were introduced by \citet{locality} in late 2006, now days they could be considered as a fundamental concepts used by every hardware manufacturer for optimization purposes. The temporal locality refers to the reuse of the same memory address within a small period of time, in other words, if a memory address is used in the present, it is more likely that the same address will be reused again in the near future. \citet{locality} defined the spatial locality as the use of very close addresses, in other words, if a memory address is used, it is more likely that nearby addresses will be used in the near future. \citet{locality-poor} also studied these principles and explained well the difference between good locality and poor locality. The ``Temporal locality'' and the ``Spatial locality'' are not the only rules that organize and influence the executed instructions and their distribution, the set of optimizations applied by compilers is also making the execution of the legitimate code more unique. The ``size optimization'' is one of these optimization where the assembly code is modified to maximize the code cache and improve the decoding time, this type of optimization involves the use of the shortest instructions, as an example, ``add eax,1000'' is used instead of ``add ebx,1000'' as it takes one byte less. Other examples of ``size optimisation'' involve using shorter addresses or exceptionally making instructions longer for the sake of alignment. \cite{optimizing} \cite{optimizing} explained hundreds of other optimization rules applied by compilers where the assembly code is modified, rearranged and placed in specific regions in order to improve its performance. The main idea of this indicator relies behind the fact that the malicious code executed during a ROP attack is different from what a normal compiler would generate as it does not obey to any rules or lows, this is because it is composed of random sequences of instructions chosen by the attacker instead of being generated by a compiler. During the execution of this malicious code, the instruction pointer points to the addresses of the gadgets instead of pointing to the program's code, it can therefore be assumed that the IP variation will be also different during a ROP attack as it reflects what has been executed. This indicator proposes monitoring the variation of this pointer and tries to identify any unusual variation or patterns that may be linked to the execution of ROP attack.

\subsection{architecture of the detection engine}

\begin{figure}[h]
\centering
\includegraphics[scale=0.3]{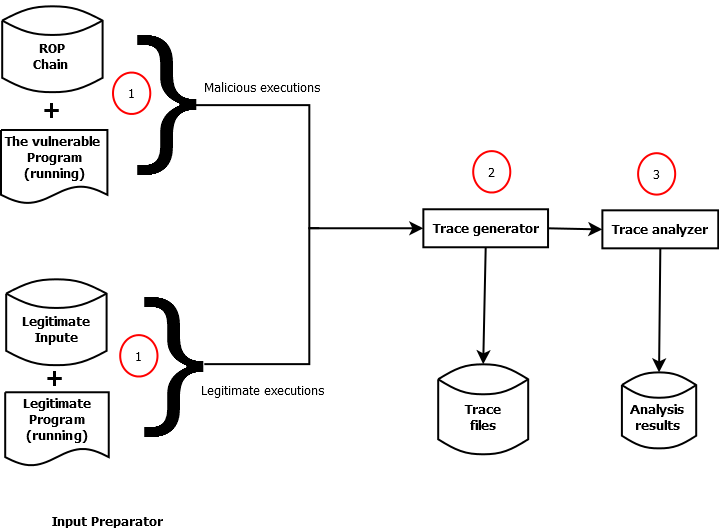}
\caption{ Global architecture of the detection engine }
\label{fig:mesure-arch}
\end{figure}

From a general perspective, the proposed architecture is composed of three modules as shown in Figure \ref{fig:mesure-arch}. Each module is responsible for performing a specific task.

As shown in Figure \ref{fig:mesure-arch}, module (1) is named ``Input preparator'', it is taking care of triggering the executable and providing the needed inputs depending on the experiment type: Running malicious executions or Running legitimate programs. In the case of studying the malicious behaviour of ROP attacks, the Input preparator executes the vulnerable program, chooses a malicious ROP chain and trigger the attack. In the case of studying the legitimate behaviour of a program, the Input preparator executes one of the programs from the listed programs, then provides it with the needed input (parameters). During the execution of the analyzed program, Module (2) takes care of performing the needed measurements, it performs a different measurement depending on the studied indicator. The results are saved into execution traces, hence the name ``Trace generator'', it is important to mention that a separated trace is generated separately for each thread running under the analyzed program. Module (3) is called the ``Trace analyzer''. As its name indicates, this module is responsible for analyzing the traces generated by the ``Trace generator''. The analysis process performed by this module consists in running the detection logic proposed by the studied indicator. The implementation of Module (2) and Module (3) (Trace generator and Trace analyzer) depends on the indicator's logic, therefore, both of them should be customized to fit the needs of each indicator. In other words, there is three different versions of Module (2) and three different versions of Module (3), however, the same implementation of Module (1) could be used for the three indicators.

\subsubsection{Trace generator \& Trace analyzer: Ret/Call parity }\label{trace-gen-1}

\begin{figure}[h]
\centering
\includegraphics[scale=0.3]{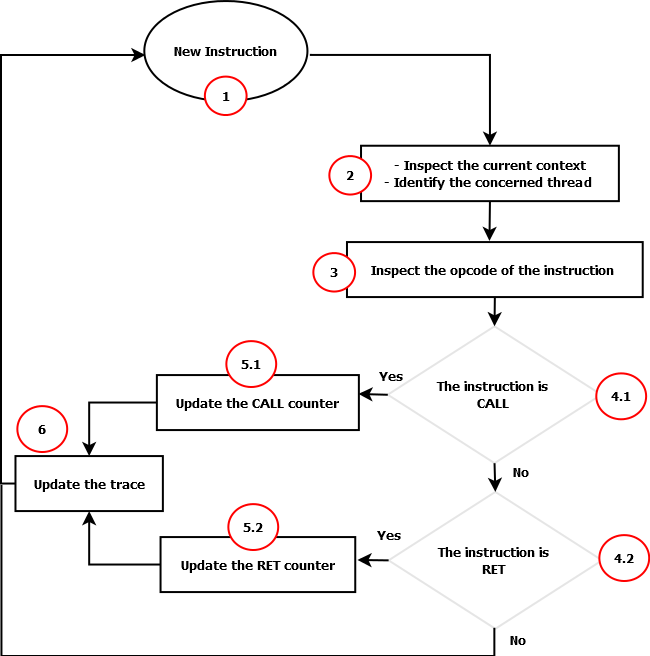}
\caption{Trace generator: Ret/Call parity }
\label{fig:tracer1}
\end{figure}

The first versions of the Trace generator and the Trace analyzer are designed for studying the first indicator. This indicator monitors the number of the executed Ret and Call at run-time and compares them to make sure that the number of the executed Call is always higher than the number of the executed Ret. The logic of the first version of the Trace generator is detailed in Figure \ref{fig:tracer1}.

As shown in Figure \ref{fig:tracer1}, step (1) consists in staying idle until the execution of a new instruction. During step (2), the module checks the current context to identify the thread in question using the unique thread ID attributed by the system. The process of identifying the thread that executes the instruction is important as a separated trace should be generated for each thread. Step (3) consists on inspecting the instruction's opcode. If the opcode corresponds to a “Call” opcode, the module increments the Call counter (5.1) and updates the trace (6). If the opcode corresponds to a Ret opcode, the module increments the Ret counter (5.2), then updates the execution trace (6). If the opcode does not match either Ret or Call, the instruction will be ignored and the module remains unresponsive until the execution of another instruction. The logic of the of Trace analyzer consists on running the verification proposed by the indicator, which could be performed in the three steps shown in Figure \ref{fig:tracer1}. Step (1) consists on waiting for a new trace update. The content of this update is inspected in step (2) to make sure that the number of the executed Call is higher than the number of the executed Ret. If not, the Trace analyzer raises an alert about a ROP attack being executed (3).

 \begin{figure}[h]
\centering
\includegraphics[scale=0.4]{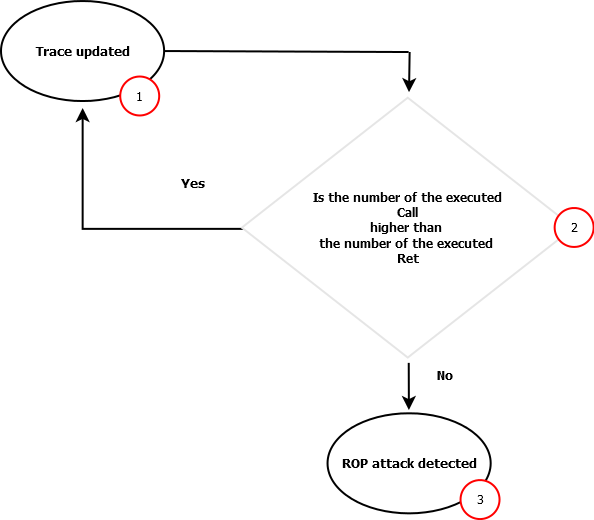}
\caption{Trace analyzer: Ret/Call parity}
\label{fig:analyzer1}
\end{figure}

\subsubsection{Trace generator \& Trace analyzer: Executed instructions between two Ret }\label{trace-gen-2}

\begin{figure}[h]
\centering
\includegraphics[scale=0.3]{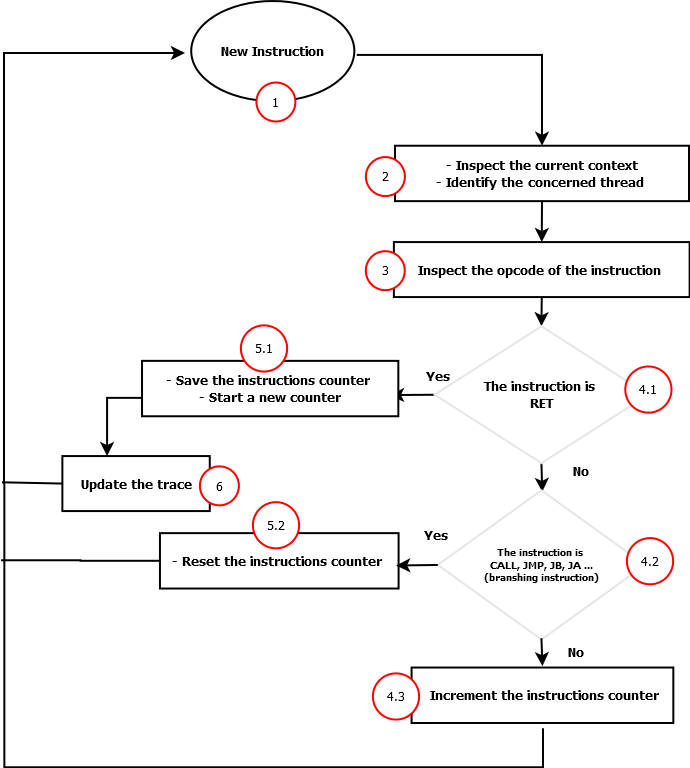}
\caption{Trace generator: Executed instructions between two Ret}
\label{fig:tracer2}
\end{figure}

The second versions of the Trace analyzer and the Trace generator are designed for the indicator that focuses on analyzing the number of the executed instructions between two consecutive Ret. The logic of the Trace generator module is detailed in Figure \ref{fig:tracer2}.

\begin{figure}[h]
\centering
\includegraphics[scale=0.3]{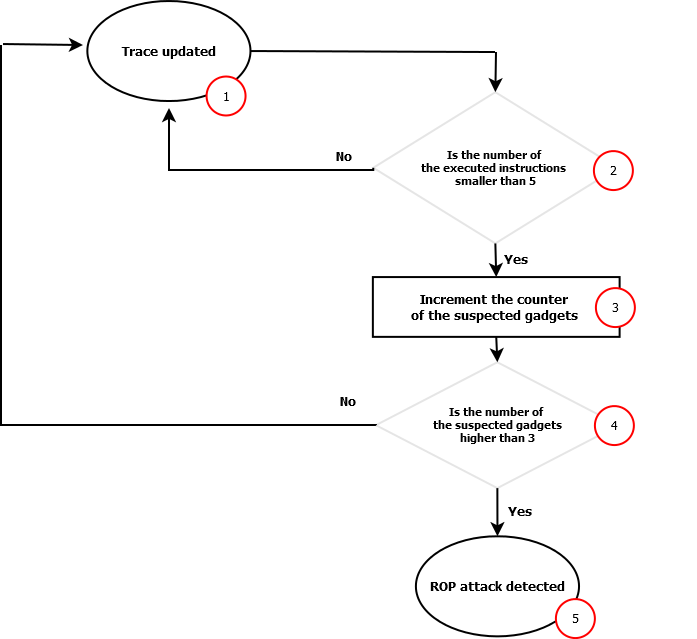}
\caption{Trace analyzer: Executed instructions between two Ret}
\label{fig:analyzer2}
\end{figure}

As shown in Figure \ref{fig:tracer2}, step (1) consists on staying idle until the execution of a new instruction, whenever an instruction is executed, the module identifies the thread in question using the unique thread ID (2). During step (3), the module inspects the instruction's opcode. If the instruction being executed is Ret (4.1), the module saves the value of the instruction counter and starts a new one (5.1). The previously saved counter is stored in the execution trace during the 6th step. If the opcode corresponds to a branching instruction (4.2), e.g., Call, Jmp, Ja, Jb or Je, the module resets the counter to zero without updating the trace (5.2). Resetting this counter each time a branching instruction is executed, is an important process as gadgets are not supposed to contain any branching instruction\cite{shacham2007geometry}. If the opcode of the executed instruction does not match either Ret or any branching instruction, the module increments the instruction counter (4.3), then remains unresponsive until the execution of the next instruction. Regarding the Trace analyzer, the main logic consists on performing the verification proposed by the second indicator. More specifically, making sure that the number of the executed instructions between two consecutive Ret is higher than five. If it is smaller than five, then it shouldn't stay that way for more than three consecutive Ret. Figure \ref{fig:analyzer2} describes the different steps performed by the trace analyzer. Step (1) consists on waiting for a new trace update. Step (2) consists on checking the updated trace and making sure that the number of the executed instructions between the last two Ret is higher than five. If not, the module increments the number of the suspected gadgets (3). During step (4), the module makes sure that the total number of the consecutive suspected gadgets is less than three. If not, it raises an alert about ROP attack being executed (5).   

\subsubsection{Trace generator: Instruction pointer variation }\label{trace-gen-3}

\begin{figure}[h]
\centering
\includegraphics[scale=0.3]{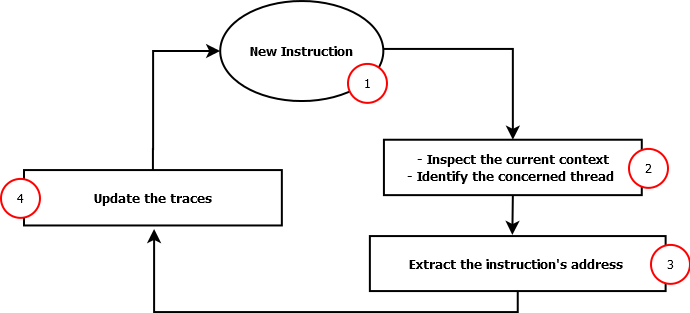}
\caption{Trace generator: Instruction pointer variation}
\label{fig:tracer3}
\end{figure}

The third version of the Trace generator is designed to study the third proposed indicator. This indicator focuses on studying the variation of the instruction pointer with the goal of identifying any unusual patterns to related the execution of ROP attacks. However, the indicator is not defining any specific patterns, therefore, it is not possible to architect a Trace analyzer module for this indicator. At this stage, the analysis process could be achieved manually, by inspecting the generated traces and trying to identify any possible patterns that may be linked to the execution of ROP attacks. This version of the Trace generator monitors the variation of the instruction pointer, extracts the address of every executed instruction then saves it in an execution trace for later analysis. This module also generates a separated trace for each thread. Figure \ref{fig:tracer3} presents the previously described logic.

 \section{THE PROOF OF CONCEPT}
 
 This part of the work aims to introduce the Proof of Concept (PoC) performed during this research work. This PoC was realized with the goal of evaluating the correctness of our hypothesis and testing the effectiveness of the three proposed indicators, the proposed measurement technique and the proposed implementation technology (Pin framework).  The first part of this section aims to introduce the reasons behind choosing the implementation technology as well as the experimental logic and setup. This part provides a detailed description of the main concept of the experiments as well as the data used during the PoC. This data represents the experimental input, and it could be categorized into two categories: the legitimate executions and the malicious executions. During the experiments, a set of tools was developed to perform some specific tasks such as the auto-generation of ROP attacks or the auto-repetition of the experiment with different data. A detailed description of these tools is also included in the first part of this chapter. The second part of this chapter takes care of summarizing and listing the results of the experimental phase. The results of each indicator were summarized in two main tables. The first table describes the results obtained after launching the 730 ROP attacks while the second table summarizes the experimental results collected after running the legitimate executions. A summary of the false positive and negative rates of each indicator was also presented in this part. The last part of this chapter aims to analyze and discuss the obtained results. The most important observations and findings are commented and detailed in this part. The false positive rate, the false negative rate, the confusion matrix, the accuracy and the error rate were also analyzed and discussed in this part. The possible causes of the false positive alerts and the false negative alerts are also discussed in this part.

\subsection{The implementation technology }\label{technology}

As discussed previously, the measurement techniques require the access to some critical run-time variables with a very high frequency (whenever an instruction is executed). Accessing such critical data with this frequency at run-time is not an easy task, hence the need for an exploitative phase before choosing the suitable technology to implement the measurement techniques. We considered three possible technologies that may be able to implement the needed tasks, and we performed an exploitative phase during which we were able to have a better idea about their performance as well as their effectiveness and ability to extract the needed data with the needed frequency. The first explored technique was the multipurpose debugger proposed by Microsoft, known as "WinDBG"\cite{windbg}. More specifically, we wanted to take advantage of the "Trace and Watch utility" presented in WinDBG. This utility can trace the execution flow and extract information at the instruction level. It consists of a multi-debugging process, during which the debugger disassembles the currently executed function and places breakpoints on each Call/Ret inside this function. If the execution hits one of these breakpoints, the debugger disassembles the called function and places the needed breakpoints. By repeating this process recursively, WinDBG becomes able to extract all the needed information (e.g. instruction's opcode and function's name). During the exploitative phase, we have discovered several limitations of this solution. One of the main limitations was the huge overhead introduced during the debugging process. Another noteworthy limitation was related to the fact that WinDBG needs the PDB files\footnote{these files are generated from the program's source code. They are usually generated during the compilation process.} of the analyzed program which is not always available. The second explored technology was a hardware-based solution. This was the new feature introduced in recent processors and known as "Last Branch Recording (LBR)"\cite{LBR}. When this feature is enabled, the CPU records the information about the last executed branching operations in special registers named "MSRs". Among the logged information we can find the destination address and the source address as well as the branching instruction (e.g. Jump, Call and Ret). This logging process is performed simultaneously while executing the program without causing any considerable slowdown. However, there still is some performance penalty for reading these MSRs. Our exploitative phase revealed that we could not rely on such technology to implement the needed measurement techniques because of the discovered limitations. The most critical limitations are related to the quantity and the quality of the provided information. More specifically, an LBR stack is limited to only 16 entries (only the last 16 branching operations are recorded) and it can be configured to only track specific types of branches. Another significant limitation is related to the fact that LBR can only be activated and accessed from kernel mode, this may not cause a problem for a research work but it can be considered as a significant disadvantage if the solution is meant to serve an end user. The third and the last explored technology was the Dynamic Binary Instrumentation (DBI). More specifically, we explored a lightweight Dynamic Binary Instrumentation framework known as ``Pin''\cite{pin}. DBI is a method used to analyze the behaviour of a binary application at run-time. This analysis is performed through the injection of some instrumentation code inside the program. The injected code executes as part of the program. Pin is known to be one of the most flexible binary instrumentation frameworks as it offers the ability to instrument the program at three different stages: Source code, Static executable level (before execution), Dynamic executable level (during the execution). Furthermore, Pin is capable of performing the instrumentation at four different levels: instruction level, the basic block level, the routine level and the image level. Our exploitative phase revealed more important advantages of Pin such as the Multiplatform support which allows the use of Pin not only in different OS’s like Windows, Linux, OSX, Android or IOS but also in different architectures like IA-32 and Intel64. However, Pin also has some drawbacks such as the considerable overhead introduced when the instruction level granularity is activated. Table \ref{tab:explorative-test} compares the explored technologies and summarizes the discovered facts during this exploitative phase. 

\begin{table}[ht]
    \centering
    \caption{Technology comparison}
    \label{tab:explorative-test}
   \includegraphics[scale=0.26]{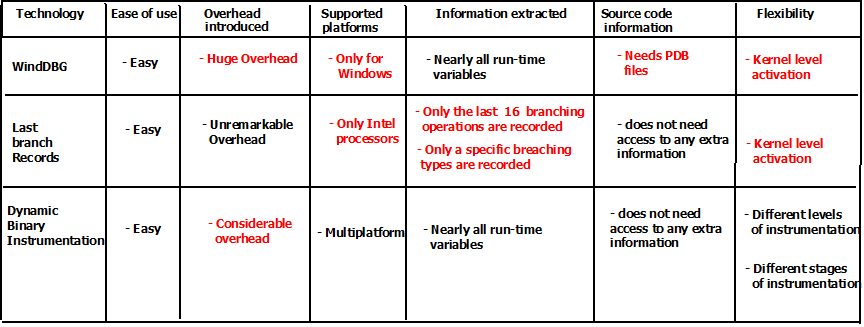}
\end{table}

As shown in Table \ref{tab:explorative-test}, it is impossible to rely on Windbg to implement our solution, first because it introduces an unacceptable overhead and second because it requires the access to the source code of the targeted program which could be unavailable in the majority of cases.  The exploitative phase revealed some critical limitations in LBR. The information extracted is limited quantitatively and qualitatively, LBR is capable of only inspecting the last 16 branches (this number can vary depending on the CPU generation but still always smaller than 32 which is in Skylake CPUs), furthermore, LBR can not inspect all types of branching operations, hence the qualitative limitation. Given these limitations, LBR can not be used to implement our solution. The facts discovered during the exploitative phase and presented in Table \ref{tab:explorative-test}, prove that the best technology to implement our proposed measurement techniques would be Pin. Pin was the most flexible technology that we explored, it offers different stages and levels of instrumentation, it does not require the access to any side information such as source code or PDB files, it allows the inspection of any run-time variable and it is a multi-platform solution that could be used in different OS’s and architectures. Based on the previously described advantages, we decided to rely on Pin framework in order to implement the main functionalists of the proposed solution.  

\section{\MakeUppercase{ Experimental logic \& Setup }}

The effectiveness of the proposed solution should be proved experimentally, this could be done by running ROP attacks against the vulnerable program and verifying whether the proposed indicators were able to detect these attacks or not. This kind of experiments can only reveal the false negative rate of the proposed solution. In order to have an idea about the false positive rate, we performed the same experiments again using legitimate executions instead of the malicious ones. Specifically, the first step of this experimental phase is to apply the measurement technique on malicious executions. The second step consists on performing the same experiments using the legitimate executions. The list of the legitimate programs/executions is described in Subsection \ref{legitim-input}. For each indicator, the experiment was repeated 1000 times using different executions, 730 are the malicious ROP attacks executed against the vulnerable program, and 270 are the legitimate programs with no ROP attacks. In order to facilitate the task of repeating these experiments, an automatization tool has been developed, this tool is described in \ref{repeat-tool}. The first versions of the Trace generator and the Trace analyzer (Subsection \ref{trace-gen-1}), were implemented and used to study the ``Ret/Call parity'' indicator. The second versions of these modules (Subsection \ref{trace-gen-2}), were implemented and used to study the indicator that focuses on studying the executed instructions between two consecutive “Ret”. The third version of the Trace generator was used to generate the execution traces for the indicator that studies the Instruction pointer variation, at this stage, there is no trace analyzer for this indicator. The implementation of these modules respects the architectures detailed in Figure \ref{fig:mesure-arch} and relies on the technology chosen in Subsection \ref{technology}.  

\subsection{The legitimate executions}\label{legitim-input}

In order to be able to calculate the false positive rate, the proposed indicators should be tested while running legitimate programs. To do so, we prepared a list of 270 executions composed of different Linux programs with different parameters. All these legitimate programs were collected from Kali2.0 OS (Debian Jessie OS, kernel version of 4.0.0). More specifically these programs were collected from the standard directory "/bin". By default, this directory usually contains the needed binaries to ensure the minimal functionality of the system. The full list of the used programs and their parameters is presented in the Appendix A. 

\subsection{The malicious executions \& The automated generation of ROP attacks }\label{auto-rop}

\begin{figure*}[ht]
\centering
\includegraphics[scale=0.56]{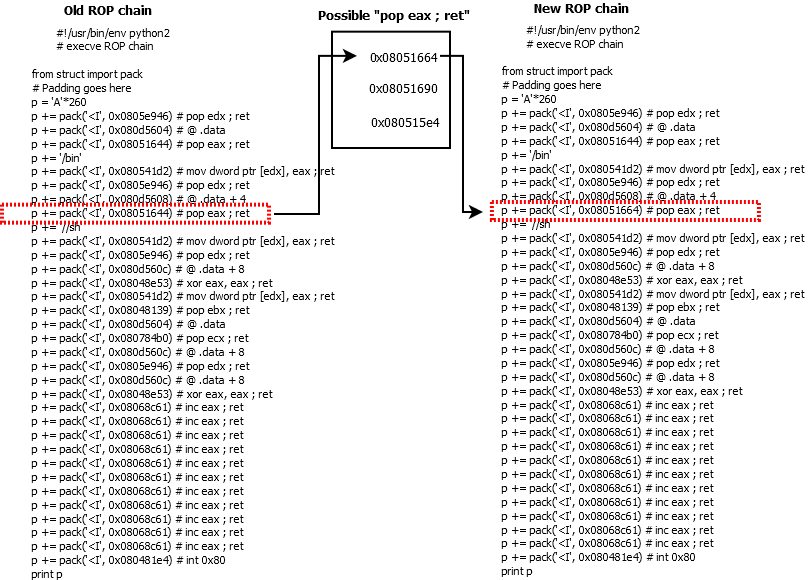}
\caption{ROP auto-generation logic }
\label{fig:rop-combination}
\end{figure*}

In order to provide enough ROP attacks during this experimental phase, we developed a Python tool named ``Auto-rop-generation-module''. This module was built based on ROPgadget \cite{ropgadget}, which is a well known software usually used by hackers/pen-testers to search for gadgets inside binaries and link them to build the ROP chain. We modified the source code of ROPgadget in order to automatically generate several ROP chains from the same binary, specifically, to generate a set of 730 ROP chains. The process of generating new ROP attacks using this tool is described in Figure \ref{fig:rop-combination}. As shown in this figure, the tool starts by using ROPgadget to generate the first ROP chain. The second step is to inspect every gadget in this chain and locate all similar gadgets. In other words, for each gadget in the main chain, we try to find the list of gadgets that can replace it. The list of the newly discovered gadgets can be used to replace the original gadgets in the main chain and build a new ROP chain. By generating all the possible combinations of these gadgets, we were able to provide enough ROP attacks for the experimental phase. Figure \ref{fig:rop-combination} describes the previously described logic used to generate these attacks. The main limitation of this process is that the generated attacks are performing the same computational operations, however, the ROP chains are completely different as they use different gadgets located in different addresses. 

It is also important to mention that we used two different sources to extract the new gadgets: the program's code (Code Segment) and the well-known C library "Libc" which is loaded by the vulnerable program. By doing so, we guarantee better diversity in the malicious sampling set. The generated attacks should be executed against a vulnerable program. Therefore, a simple vulnerable C program has been proposed, this program performs some random computational operations using some C functions. Among these functions, there is the well known "memcpy", which copies the input to a memory destination without doing bounds checking. First, memcpy reads the input, and then it saves it into a destination buffer. The typical exploitation technique consists on providing a large input for this function, larger than the destination buffer size. The large input will exceed the buffer’s bounds and overwrite adjacent information such as function’s return address. 

\subsection{The automatisation tool }\label{repeat-tool}

\begin{figure}[ht]
\centering
\includegraphics[scale=0.28]{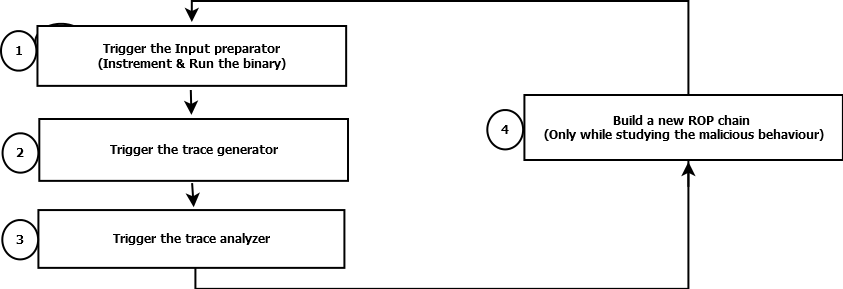}
\caption{Automatisation module}
\label{fig:automatisaztion}
\end{figure}

This tool was built in order to facilitate the task of repeating these experiments with different inputs. It allows the coordination between the four main modules: the Input preparator, the Trace generator, the Trace analyzer and the Auto-rop-generation-module. The tool was built using Python 2.7. Figure \ref{fig:automatisaztion} illustrates the logic of this module.

\section{\MakeUppercase{ Experimental Results }}\label{results-exprements}

\subsection{Results of the first indicator: Ret/Call parity analysis}

\begin{figure}[ht]
\centering
\includegraphics[scale=0.3]{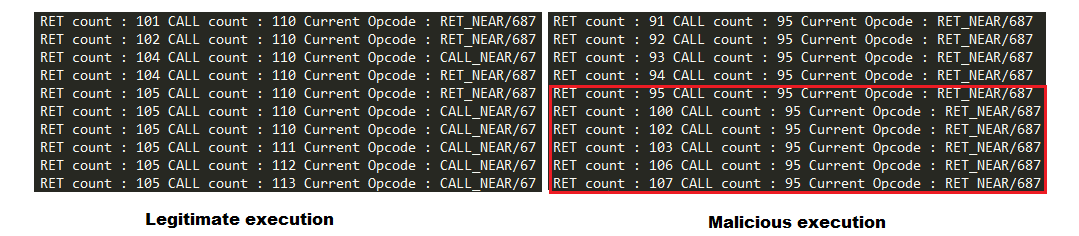}
\caption{Traces screen-shot}
\label{fig:ret-call-trace}
\end{figure}

As discussed, the experiment was performed against two types of executions: Malicious and legitimate. The trace generator took care of generating the traces and the trace analyzer analyzed them in order to decide whether a ROP attack is being executed or not. The left side of Figure \ref{fig:ret-call-trace} shows an example of a legitimate trace generated by our trace generator. The same trace was regenerated while performing a ROP attack (the right side of Figure \ref{fig:ret-call-trace}), as we can see in the right side of Figure \ref{fig:ret-call-trace}, when a ROP attack is triggered, the number of the executed Ret is becoming higher than the number of the executed Call, for that reason, it was flagged as malicious by our trace analyzer module. 

The first part consisted on repeating the experiment against 730 ROP attack in different conditions: different Gadget sources, DEP on/off and ASLR on/off. During this phase, 30 ROP attack were generated using Gadgets from the program's code (The code segment), and 700 attacks were constructed using gadgets from Libc. The overhead introduced by our tools was also collected for further analysis. The results of this experimental phase are summarized in Table \ref{tab:ioc1-results-tabl}. 

\begin{table}[ht]
    \centering
     \caption{Results of the first indicator: Malicious executions}
   \label{tab:ioc1-results-tabl}
\begin{tabular}{ |p{2cm}||p{1cm}|p{1cm}|p{1cm}|p{1cm}|p{1cm}|p{1cm}|  }
\hline
Gadget Source & Samples Number & DEP & ASLR &  Detected  \\
 \hline
* Code segment (the program code ) &  30  &  Disabled & Disabled  & 100\%  \\
* Libc  &  700  &   Disabled & Disabled & 100\% \\
 \hline
 Code segment (the program code ) &  30  &  Enabled  & Disabled & 100\%  \\
 Libc  &  700  &  Enabled  & Disabled & 100\%  \\
 \hline
\hline 
 Code segment (the program code ) &  15  &  Disabled & Enabled  & 100\% \\
 Libc  &  50  & Disabled & Enabled   & 100\%  \\
 \hline
  Code segment (the program code ) &  15 &  Enabled & Enabled  & 100\%  \\
 Libc  &  50  & Enabled & Enabled   & 100\% \\
 \hline
  \hline
\end{tabular} \\

\end{table}

The second part of this Proof of Concept consisted on performing the same experiment using the legitimate executions instead of the vulnerable program and the ROP attacks. Unlike the previous part, these experiments were performed only with DEP activated and ASLR deactivated as the main goal was to have an idea about the false positive rate. The results of these tests are summarized in Table \ref{tab:ioc1-results-tabl-legitimate}.  

\begin{table}[ht]
    \centering
     \caption{Results of the first indicator: Legitimate executions }
   \label{tab:ioc1-results-tabl-legitimate}
\begin{tabular}{ |p{3cm}|p{3cm}|p{3cm } | }
\hline

 Samples number & Detected  \\
 \hline

270 & 1.1\% \\
 \hline

\end{tabular} \\

\end{table}

The results presented in Tables \ref{tab:ioc1-results-tabl} and \ref{tab:ioc1-results-tabl-legitimate} can be used to construct the confusion matrix which summarizes the false negative rate (FN) and true positive rate (TP). This matrix is presented in Table \ref{tab:confusion-matrix1}.

\begin{table}[ht]
    \centering
    \caption{Confusion Matrix: First Indicator}
    \label{tab:confusion-matrix1}
\begin{tabular}{ p{1.5cm}|p{1.5cm}|p{1.5cm }| p{1cm } }
 & Detected as Legitimate&Detected as Malicious&   \\
 \hline
 Actually Legitimate & TN = 267  & FP = 3 & N =  270  \\
\hline
Actually Malicious  & FN = 0 & TP = 730 & P= 730 \\
\hline
  &  267  & 733 & - \\
\end{tabular} \\
\end{table}

\subsection{Results of the second indicator: Instructions number between two Ret}

\begin{figure}[ht]
\centering
\includegraphics[scale=0.4]{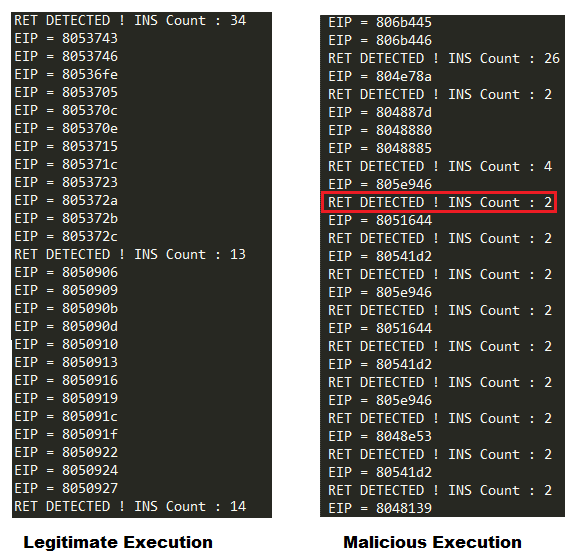}
\caption{Traces screen-shot}
\label{fig:inst-count-trace}
\end{figure}

In order to prove the effectiveness of the second proposed indicator, we followed the same experimental logic described previously, except that we used the second version of the trace generator and the trace analyzer. The left side of Figure \ref{fig:inst-count-trace} shows an example of a legitimate trace generated by our trace generator, while the right side represents a malicious trace flagged as ROP attack by our trace analyzer. As shown in the malicious trace, during a ROP attack, the number of the instructions executed between two Ret becomes small (smaller than five) when the attack is triggered, this behaviour is captured by our trace analyzer module and interpreted as ROP attack. The trace generator counts the second executed Ret among the executed instructions, but the trace analyzer takes this into consideration while performing the detection. The address of every executed instruction was also collected, this information could be useful for further analysis.

This experiment was repeated 1000 times against different executions. The same 730 ROP attacks and the same 270 legitimate programs from the previous experiments were used during this experimental phase. The malicious executions were also performed in different conditions (different Gadget sources, DEP on/off and ASLR on/off). The results of the malicious executions are summarized in Table \ref{tab:ioc2-res-malicous}. 

\begin{table}[ht]
    \centering
     \caption{Results of the second indicator: Malicious execution}
    \label{tab:ioc2-res-malicous}
\begin{tabular}{ |p{2cm}||p{1cm}|p{1cm}|p{1cm}|p{1cm}|p{1.5cm}|  }
\hline
Gadget Source & Samples Number & DEP & ASLR &  Detected\\
 \hline
* Code segment (the program code ) &  30  &  Disabled & Disabled  & 100\%  \\
* Libc  &  700  &   Disabled & Disabled & 100\% \\
 \hline
 Code segment (the program code ) &  30  &  Enabled  & Disabled & 100\%  \\
 Libc  &  700  &  Enabled  & Disabled & 100\% \\
 \hline
\hline 
 
 Code segment (the program code ) &  15  &  Disabled & Enabled  & 100\%  \\
 Libc  &  50  & Disabled & Enabled   & 100\%  \\
 
 \hline
 
  Code segment (the program code ) &  15 &  Enabled & Enabled  & 100\% \\
 Libc  &  50  & Enabled & Enabled   & 100\%  \\
 \hline
 
  \hline

\end{tabular} \\

\end{table}

The second phase consisted on testing the indicator against the non-malicious executions, this kind of tests allows the calculation of the false positive rate. The results of this second phase are summarized in Table \ref{tab:ioc2-results-tabl-legitimate}: 

\begin{table}[ht]
    \centering
    \caption{Results of the second indicator: Legitimate executions}
    \label{tab:ioc2-results-tabl-legitimate}

\begin{center}

\begin{tabular}{ |p{3cm}|p{3cm}|p{3cm }| }
\hline

 Samples number & Detected    \\
 \hline

270 & 1.8\% \\
 \hline
\end{tabular} \\
\end{center}
\end{table}

The confusion matrix of this indicator is presented \ref{tab:confusion-matrix2}, it was constructed using the results presented in Tables \ref{tab:ioc2-res-malicous} and \ref{tab:ioc2-results-tabl-legitimate}. 

\begin{table}[ht]
    \centering
    \caption{Confusion Matrix: Second Indicator}
    \label{tab:confusion-matrix2}
  
\begin{center}
    
\begin{tabular}{  p{1.5cm}|p{1.5cm}|p{1.5cm }| p{1cm }  }

 & Detected as Legitimate&Detected as Malicious&   \\
  
 \hline

 Actually Legitimate & TN = 265  & FP = 5 & N =  270  \\

\hline

Actually Malicious  & FN = 0 & TP = 730 & P= 730 \\

\hline
  &  265  & 735 & - \\

\end{tabular} \\
\end{center}
\end{table}

\subsection{Results of the third indicator: Analyzing the IP/EIP variation }

\begin{figure}[ht]
\centering
\includegraphics[scale=0.8]{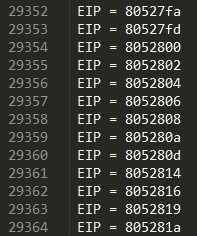}
\caption{Trace screen-shot}
\label{fig:eip-var-trace2}
\end{figure}

Unlike the previous indicators, at this stage, this one is not proposing any exact patterns to be recognized, therefore, it was not possible to perform a process of auto-detection using a trace analyzer. However, the same experiments were performed on this indicator, and the execution traces were generated using the third version of the trace generator. These traces were analyzed manually with the goal of proving the existence of new patterns appearing in the IP variation when a ROP attack is being executed. Figure \ref{fig:eip-var-trace2} represents an example from the generated traces, it describes the different values assigned to the instruction pointer (IP or EIP) during the execution. In other words, the trace contains the addresses of the executed instructions.

\begin{center}
\begin{figure}[ht]
\includegraphics[scale=0.28]{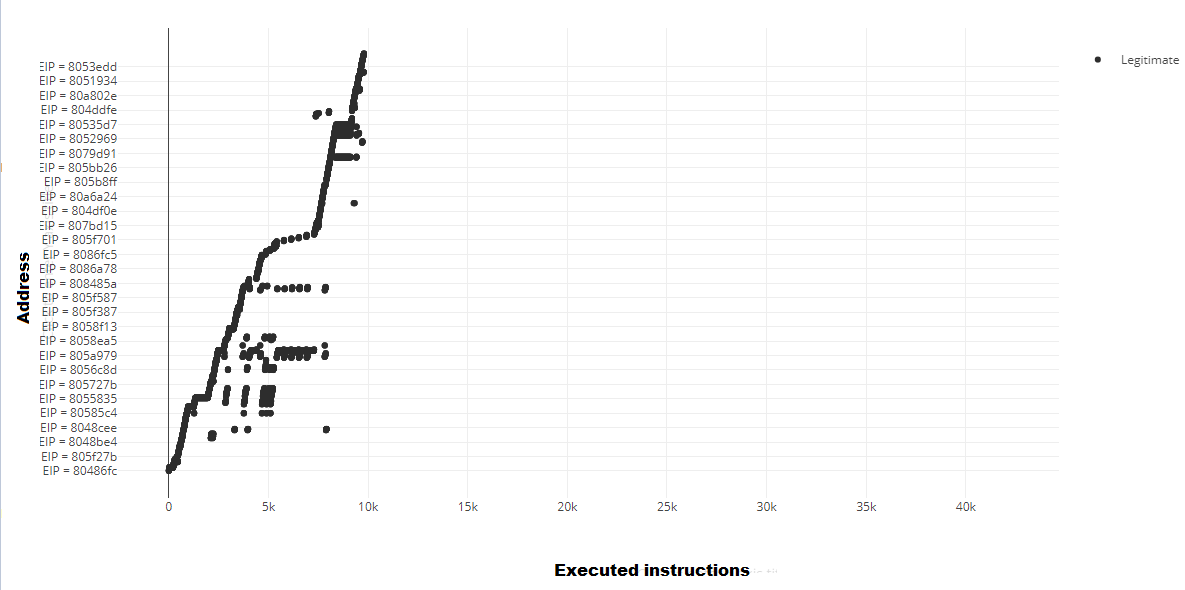}
\caption{ EIP variation during the legitimate execution }
\label{fig:legitimate-eip}
\end{figure}
\end{center}

\begin{center}
\begin{figure}[ht]
\includegraphics[scale=0.28]{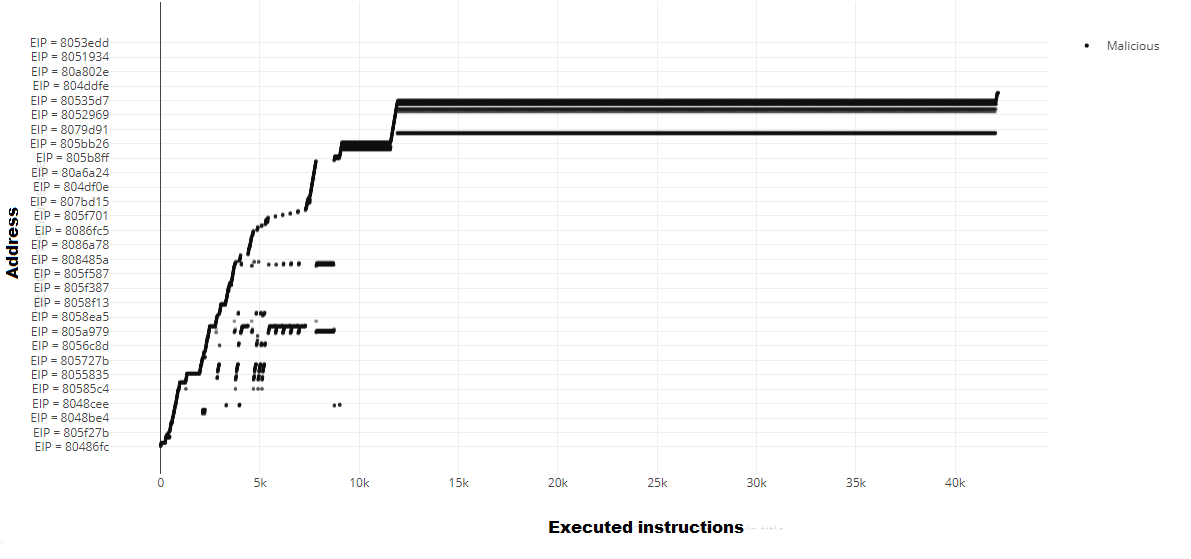}
\caption{ EIP variation during the malicious execution }
\label{fig:vuln-eip}
\end{figure}
\end{center}

The manual analysis of these traces revealed that the EIP variation is indeed witnessing the apparition of new patterns during a ROP attack. In order to make the analysis easier, some examples of the generated traces were plotted, Figure \ref{fig:legitimate-eip} and Figure \ref{fig:vuln-eip} are presenting two different plotted executions of the same program. The first one is the legitimate execution and the second one is the malicious one. The new patterns that appear during the malicious executions and their causes will be discussed and analyzed in the next sections.

\section{\MakeUppercase{ discussion and Limitations }}

\subsection{Discussing the results of the first indicator: Ret/Call parity analysis}

\begin{figure}[ht]
\centering
\includegraphics[scale=0.25]{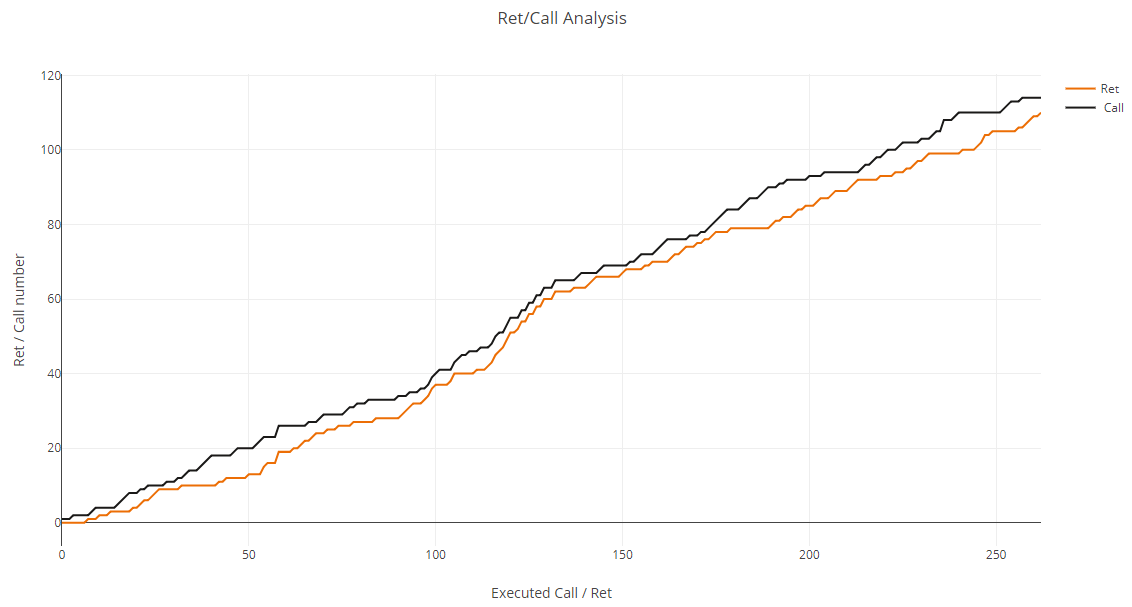}
\caption{Legitimate Ret/Call }
\label{fig:ret-call-legitimate}
\end{figure}

\begin{figure}[h]
\centering
\includegraphics[scale=0.2]{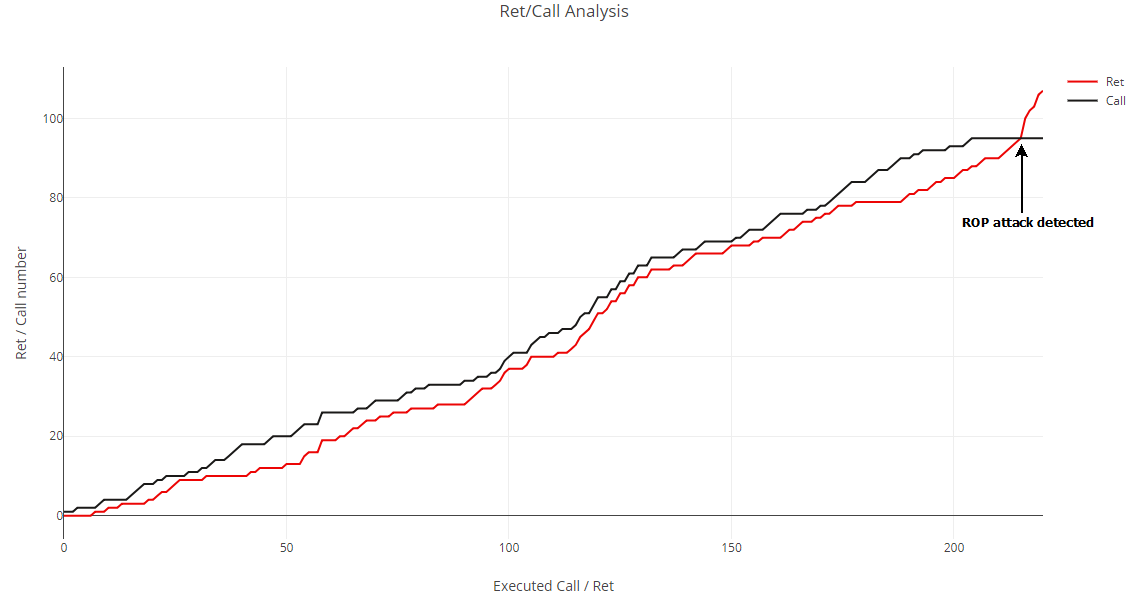}
\caption{Malicous Ret/Call }
\label{fig:ioc1-res}
\end{figure}

 After performing the experimental phase and taking a closer look at the collected results, it becomes clear that this indicator has an excellent effectiveness in detecting the existence of ROP attacks as there was no false negative alerts. However, the generalisability of these results is subject to certain limitations. In order to generalize these results, the same experiment has to be performed against a larger sampling set that includes a variety of vulnerable programs and a wider ROP chain set. In order to facilitate the analysis of the obtained results, some examples of the generated traces were plotted, figures \ref{fig:ioc1-res} and \ref{fig:ret-call-legitimate} describe two plotted traces of the same program, the first figure represents the trace of the legitimate execution while the second one represents the trace of the malicious execution obtained after running a ROP attack. As shown in Figure \ref{fig:ret-call-legitimate}, during a legitimate execution, the number of the executed Call is always higher than the number of the executed Ret. 
 
 As shown in Figure \ref{fig:ioc1-res}, when the exploit is triggered, the number of the executed Ret starts to increase till exceeding the number of the executed Calls. This behaviour reflects exactly the logic used by our indicator to identify these attacks. Another important observation was about the fact that the number of the executed Calls keeps the same value during the attack, this fact is logical as gadgets are not supposed to contain a Call instruction. This fact was not expected, but it is considered as one of the most important observations and a possible future enhancement for this indicator in future works. Despite the good results, it is not possible to guarantee that this indicator will always keep the same false negative rate for any sampling set. In order to be able to draw more generalized conclusions, a wider sampling set should be used in future works, this sampling set should contain more vulnerable programs tested with real word ROP chain instead of the auto-generated attacks. However, the concrete conclusion that can be drawn from the present study is about the excellent effectiveness of this indicator in detecting ROP attacks. There are two possible causes of having a false negative alert triggered by this indicator: 

\begin{itemize}
    
\item  Exceptions are the first possible cause: Exceptions force the program to jump and execute the exception code, this jump will prevent the normal execution from hitting/executing the Ret associated with the previously executed Call. If this behaviour is repeated more than the number of the gadgets used in the ROP chain, then this indicator will fail in detecting the attack. 

\item  Go-to statements are the second possible cause: Go-to statements can force the program to jump to a new code location without executing the Ret instruction associated with the previously executed Call. If such behaviour is repeated more than the number of the gadgets in the ROP chain, then this indicator will fail in detecting the attack.
\end{itemize}

The results presented in Table \ref{tab:ioc1-results-tabl-legitimate} show that this indicator suffers from a false positive rate equal to 1.1\%, three executions were flagged as malicious, but they are not. One of the possible causes of these false positive alerts is when the program jumps to the beginning of a function instead of using the Call instruction. In other words, when the Call instruction associated with the Ret is never executed, in order to confirm this, a further reverse engineering process should be performed against the flagged programs. The confusion matrix of this indicator (previously presented in Table \ref{tab:confusion-matrix1}) permits the calculation of the ``Accuracy'' and the ``Error Rate'', these rates describe how often the results are correct and how often they are wrong. The "Accuracy" of this indicator is 0.997, and it could be calculated using the Equation \ref{auucurancy}. Having an accuracy equal to 0.997 means that the results of this indicator are correct at 99.7\%. This accuracy could be considered as very good as the best of the 56 VirusTotal AV engines has an accuracy 99.4\%\cite*{accurancy-best}. The Error rate is equal to 0.003 and could be calculated using the Equation \ref{error-rate}. This value reflects the fact that the results are wrong in 0.3\% of the cases.

\begin{equation}
    Accuracy = (TP+TN)/P+N 
    \label{auucurancy}
\end{equation}
  
\begin{equation}
    Error Rate = (FP+FN)/P+N 
    \label{error-rate}
\end{equation}

\subsection{Discussing the results of the second indicator: Instructions number between two Ret}

\begin{center}
\begin{figure}[h]
\centering
\includegraphics[scale=0.3]{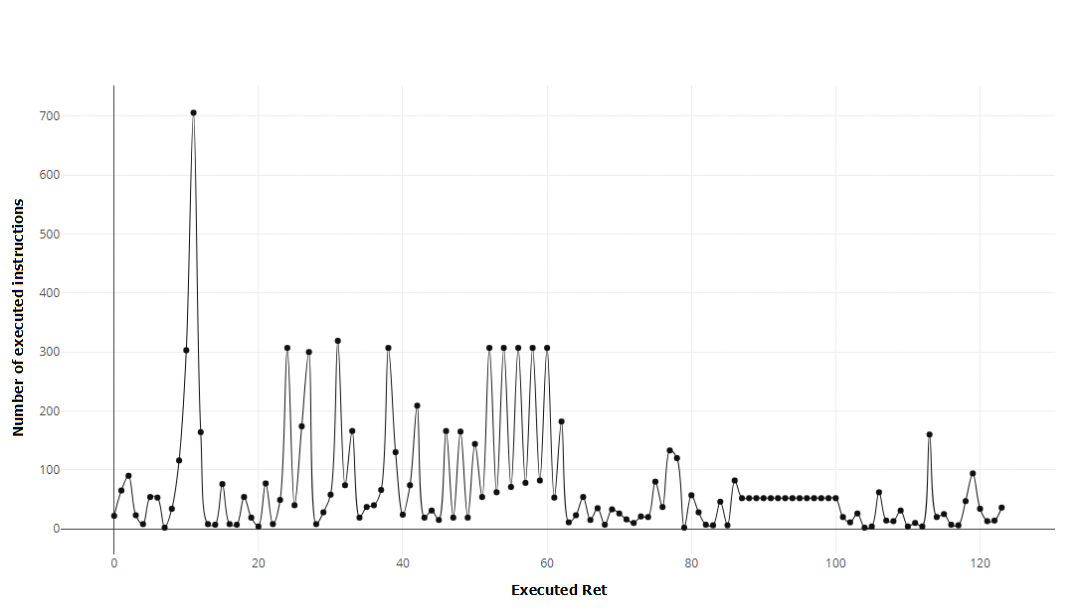}
\caption{ Number of executed instructions between two Ret (Legitimate execution)}
\label{fig:ioc2-res}
\end{figure}
\end{center}

\begin{center}
\begin{figure}[h]
\centering
\includegraphics[scale=0.3]{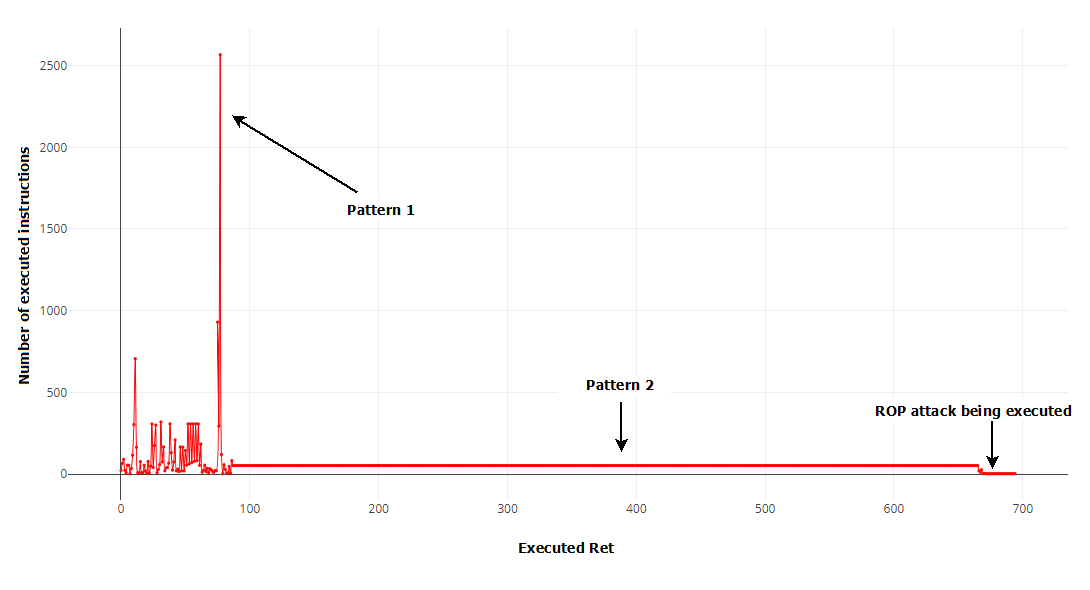}
\caption{ Number of executed instructions between two Ret (Malicious execution)}
\label{fig:ioc2-res2}
\end{figure}
\end{center}

\begin{center}
\begin{figure}[h]
\centering
\includegraphics[scale=0.3]{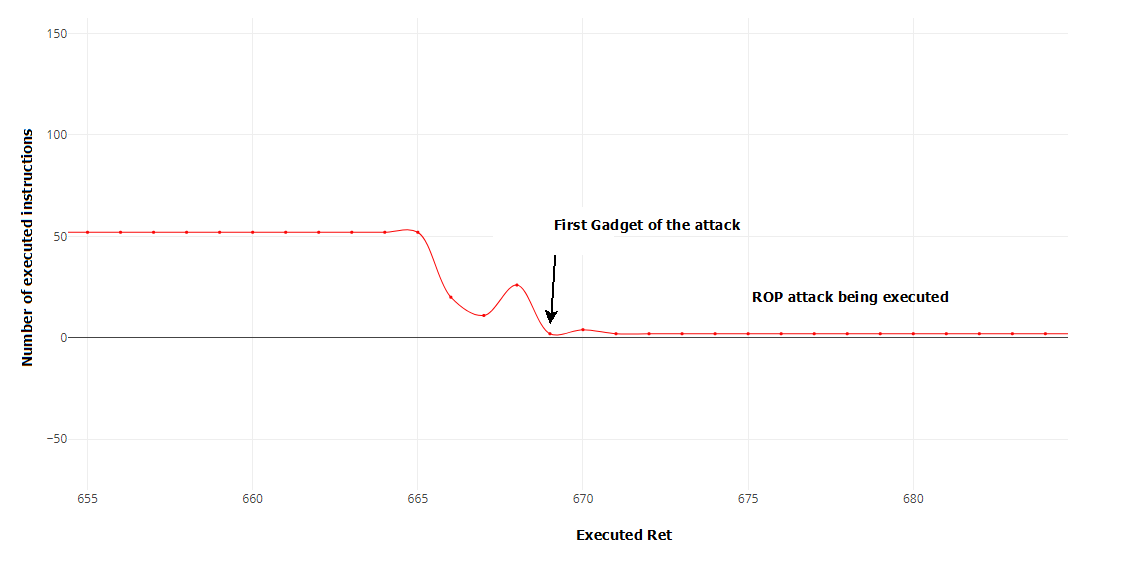}
\caption{ Zoomed ROP chain}
\label{fig:ioc2-res2-zoom}
\end{figure}
\end{center}

As shown in Table \ref{tab:ioc2-res-malicous}, the false negative rate is null. However, in order to draw a generalized conclusion about this rate, further experiments must be performed against a wider sampling set that includes the use of different vulnerable programs and more diversified ROP chains. Regarding the non-malicious executions, 1.8\% of them were flagged as malicious but they are not. The causes behind these false positive alerts could be discovered by analyzing both the execution traces generated by our tool and the dissembled programs. One of the possible causes could be related to the execution of a recursive loop containing a small number of instructions (ex. incremental loop). In order to improve the false positive rate in future work, an extra step of verification could be added to this indicator, the verification consists on checking if there are any patterns in the executed instructions, if so, then this could be a loop as it repeats the same instructions. Some examples of the generated traces were plotted with the goal of having a better view of the collected data, Figure \ref{fig:ioc2-res} describes an example of a plotted trace, this trace represents a legitimate execution of our vulnerable program. Figure \ref{fig:ioc2-res2} represents the execution of the same vulnerable program while running the ROP attack. This figure demonstrates clearly that when the ROP chain is triggered, the number of the instructions between two Ret falls suddenly and becomes very small. Figure \ref{fig:ioc2-res2-zoom} is showing better this behaviour after re-scaling the graph. First, the plotted malicious execution starts with the same variation as the legitimate one, then it witnesses the apparition of a new pattern (Pattern 1). This pattern represents an unusual behaviour generated by ROP attacks, during which the number of the instructions executed between two consecutive Ret is raising highly before becoming small again. This pattern was considered as an ``artifact'' that can help to identify ROP attacks and could be added in future work to improve the detection rate of this indicator. After inspecting and analyzing this unexpected behaviour, we found that it is generated by the large input used to trigger the overflow in order to hijack the control flow. In other words, it is generated by the memcpy function when it is looping and copying this data. The second pattern is still unexplained but we firmly believe that it is linked to some functions called by memcpy.

The Accuracy and the Error rate of this indicator were also calculated using the confusion matrix presented in Table \ref{tab:confusion-matrix2}. The Accuracy is equal to 0.995 which means that the indicator is correct in 99.5\% of the cases, this could be considered as good results compared to other public detectors\cite{accurancy-best}. The Error rate is 0.005, this rate reflects the fact that the results could be wrong in 0.5\% of the cases, which is a very acceptable rate\cite{accurancy-best}.   

\subsection{Discussing the results of the thirds indicator: EIP variation}

\begin{figure}[h]
\centering
\includegraphics[scale=0.25]{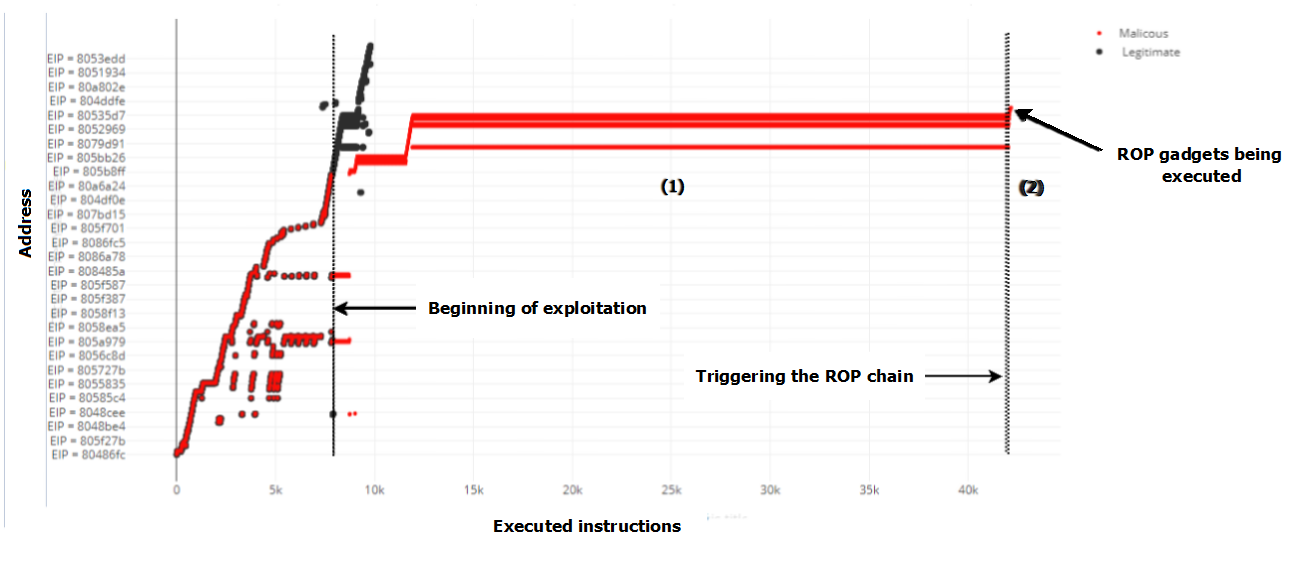}
\caption{Comparing the malicious EIP variation with the legitimate variation}
\label{fig:eip-var-trace}
\end{figure}

Figure \ref{fig:eip-var-trace} represents a comparison of the two figures \ref{fig:legitimate-eip} and \ref{fig:vuln-eip}, the red graph represents the malicious execution while the black graph represents the legitimate execution. As shown in Figure \ref{fig:eip-var-trace}, the malicious and the legitimate variation start with the exact same patterns, this behaviour is observed from instruction 0 to around instruction 7K. 

 When the exploitation process starts around instruction 7k, the malicious EIP variation witnesses the apparition of new patterns. More specifically two new clear patterns are being created:

\begin{itemize}

    \item Pattern (1): Small variations in the EIP that keeps repeating and jumping from/to the same addresses. This pattern starts near instruction 7K and ends near instruction 42k. This variation represents a big loop being executed as the same instructions are being executed several times. 
    \item Pattern (2): A very small random variation at the end near instruction 43k, this variation is represented by the scattered red points at the top right side of the figure. 
    
\end{itemize}

An in-depth inspection was conducted to reveal the causes behind the apparition of the two patterns. It has been discovered that the first pattern is created by the memcpy function, which is the vulnerable function being exploited to hijack the control flow. More specifically, during the exploitation phase, we introduced 260 characters + Malicious Payload as an input for the memcpy in order to trigger the overflow, memcpy is looping to read and copy the 260 characters which explain the apparition of this pattern. It is also important to mention that the same behaviour was observed differently when we tested the second indicator. The deep inspection of the collected traces revealed that the second pattern is caused by the execution of the ROP chain itself, it is describing the EIP jumping from a gadget to another and moving from instruction to another in the gadgets. 

\section{CONCLUSION}

The findings in this study are subject to at least three main limitations. The first and the most important limitation is regarding the third indicator. Although the study has successfully demonstrated that the first two indicators are fully capable of identifying ROP attacks
at run-time, it has certain limitations regarding proving the effectiveness of the third proposed indicator. This indicator still not specific enough and needs a concrete specification of the IP patterns that should be used to identify ROP attacks. However, the analysis of
the collected traces was helpful as it revealed some possible patterns. Future work should focus on studying these possible patterns and analyzing the collected traces deeply in order to be able to extract the exact IP variation patterns that could be used to identify ROP attacks.

The second major limitation is regarding the auto-generated ROP chains used during the experimental phase. The proposed technique to generate ROP attacks is building a set of different ROP chains but unfortunately they are similar from a computational point of view.
In other words, the generated ROP chains are performing the same computational operations. We think that such similarity may poison the results obtained during the experimental phase. Future research should therefore concentrate on studying the performance of the proposed
indicators against real word ROP attacks instead of using auto-generated attacks.

The third limitation of this study lies in the fact that the experimental phase was performed using only one vulnerable program (developed in the lab). In other words, the study was not able to provide a good diversity of the set of vulnerable programs. Therefore, caution must be applied, as the results (e.g. the detecting rate) might not be the same if a wider set of vulnerable programs is considered. Therefore, further experimental investigations are needed to reevaluate the experimental results while considering a wider set of vulnerable programs. It is also important to point to an important issue that was not addressed in this study, this issue is about inspecting the causes of the false positive alerts observed during the experimental phase. It was not possible to investigate the real causes of these alerts as it is considered out of the scope of this research work, however, it is recommended that further
research be undertaken to investigate these causes as this would help us to first establish a greater degree of accuracy and second improve performance of the proposed solution.

\addtolength{\textheight}{-12cm}   





\bibliographystyle{IEEEtranN}
\bibliography{references}

\begin{thebibliography}{43}
\providecommand{\natexlab}[1]{#1}
\providecommand{\url}[1]{#1}
\csname url@samestyle\endcsname
\providecommand{\newblock}{\relax}
\providecommand{\bibinfo}[2]{#2}
\providecommand{\BIBentrySTDinterwordspacing}{\spaceskip=0pt\relax}
\providecommand{\BIBentryALTinterwordstretchfactor}{4}
\providecommand{\BIBentryALTinterwordspacing}{\spaceskip=\fontdimen2\font plus
\BIBentryALTinterwordstretchfactor\fontdimen3\font minus
  \fontdimen4\font\relax}
\providecommand{\BIBforeignlanguage}[2]{{%
\expandafter\ifx\csname l@#1\endcsname\relax
\typeout{** WARNING: IEEEtranN.bst: No hyphenation pattern has been}%
\typeout{** loaded for the language `#1'. Using the pattern for}%
\typeout{** the default language instead.}%
\else
\language=\csname l@#1\endcsname
\fi
#2}}
\providecommand{\BIBdecl}{\relax}
\BIBdecl

\bibitem[Chen et~al.(2005)Chen, Xu, Nakka, Kalbarczyk, and Iyer]{memcorruption}
S.~Chen, J.~Xu, N.~Nakka, Z.~Kalbarczyk, and R.~K. Iyer, ``Defeating memory
  corruption attacks via pointer taintedness detection,'' in \emph{Dependable
  Systems and Networks, 2005. DSN 2005. Proceedings. International Conference
  on}.\hskip 1em plus 0.5em minus 0.4em\relax IEEE, 2005, pp. 378--387.

\bibitem[Tawalbeh et~al.(2017)Tawalbeh, Houssain, and Al-Somani]{sidechanel}
L.~Tawalbeh, H.~Houssain, and T.~Al-Somani, ``Review of side channel attacks
  and countermeasures on ecc, rsa, and aes cryptosystems,'' vol.~6, 04 2017.

\bibitem[Andersen and Abella(2004)]{DEP}
S.~Andersen and V.~Abella, ``Data execution prevention. changes to
  functionality in microsoft windows xp service pack 2, part 3: Memory
  protection technologies,'' 2004.

\bibitem[Bletsch(2011)]{codereuseattacks}
T.~Bletsch, \emph{Code-reuse attacks: new frontiers and defenses}.\hskip 1em
  plus 0.5em minus 0.4em\relax North Carolina State University, 2011.

\bibitem[Larsen and Sadeghi(2018)]{codereuseattacks2}
P.~Larsen and A.-R. Sadeghi, \emph{The Continuing Arms Race: Code-Reuse Attacks
  and Defenses}.\hskip 1em plus 0.5em minus 0.4em\relax Morgan \& Claypool,
  2018.

\bibitem[Shacham(2007)]{shacham2007geometry}
H.~Shacham, ``The geometry of innocent flesh on the bone: Return-into-libc
  without function calls (on the x86),'' in \emph{Proceedings of the 14th ACM
  conference on Computer and communications security}.\hskip 1em plus 0.5em
  minus 0.4em\relax ACM, 2007, pp. 552--561.

\bibitem[Irvine and Das(2011)]{x86assembly}
K.~R. Irvine and L.~B. Das, \emph{Assembly language for x86 processors}.\hskip
  1em plus 0.5em minus 0.4em\relax Prentice Hall, 2011.

\bibitem[Carlini and Wagner(2014)]{ropdanger}
N.~Carlini and D.~Wagner, ``Rop is still dangerous: Breaking modern defenses.''
  in \emph{USENIX Security Symposium}, 2014, pp. 385--399.

\bibitem[Homescu et~al.(2012)Homescu, Stewart, Larsen, Brunthaler, and
  Franz]{turingrop}
A.~Homescu, M.~Stewart, P.~Larsen, S.~Brunthaler, and M.~Franz, ``Microgadgets:
  size does matter in turingcomplete return-oriented programming,'' in
  \emph{Proceedings of the 6th USENIX conference on Offensive
  Technologies}.\hskip 1em plus 0.5em minus 0.4em\relax USENIX Association,
  2012, pp. 7--7.

\bibitem[Bletsch et~al.(2011{\natexlab{a}})Bletsch, Jiang, Freeh, and
  Liang]{jop}
T.~Bletsch, X.~Jiang, V.~W. Freeh, and Z.~Liang, ``Jump-oriented programming: a
  new class of code-reuse attack,'' in \emph{Proceedings of the 6th ACM
  Symposium on Information, Computer and Communications Security}.\hskip 1em
  plus 0.5em minus 0.4em\relax ACM, 2011, pp. 30--40.

\bibitem[Team(2003)]{aslr}
P.~Team, ``Pax address space layout randomization (aslr), 2003,'' \emph{URL:
  https://pax. grsecurity. net/docs/aslr. txt (visited on 12/14/2017)},
  vol.~54, 2003.

\bibitem[Infosec(2014)]{ret-to-plt}
Infosec, ``Bypassing non-executable-stack during exploitation using
  return-to-libc,'' 2014.

\bibitem[Gruss et~al.(2016)Gruss, Maurice, Fogh, Lipp, and Mangard]{AnC2017}
D.~Gruss, C.~Maurice, A.~Fogh, M.~Lipp, and S.~Mangard, ``Prefetch side-channel
  attacks: Bypassing smap and kernel aslr,'' in \emph{Proceedings of the 2016
  ACM SIGSAC conference on computer and communications security}.\hskip 1em
  plus 0.5em minus 0.4em\relax ACM, 2016, pp. 368--379.

\bibitem[Evtyushkin et~al.(2016)Evtyushkin, Ponomarev, and
  Abu-Ghazaleh]{aslr-bypass2}
D.~Evtyushkin, D.~Ponomarev, and N.~Abu-Ghazaleh, ``Jump over aslr: Attacking
  branch predictors to bypass aslr,'' in \emph{Microarchitecture (MICRO), 2016
  49th Annual IEEE/ACM International Symposium on}.\hskip 1em plus 0.5em minus
  0.4em\relax IEEE, 2016, pp. 1--13.

\bibitem[Christey(2007)]{cve}
S.~Christey, ``Unforgivable vulnerabilities,'' \emph{Black Hat Briefings},
  vol.~13, p.~17, 2007.

\bibitem[Kania and Costello(2018)]{clandestine-wolf}
E.~B. Kania and J.~K. Costello, ``The strategic support force and the future of
  chinese information operations,'' \emph{The Cyber Defense Review}, vol.~3,
  no.~1, pp. 105--122, 2018.

\bibitem[Thakur(2017)]{greedywonk}
K.~Thakur, ``Hybrid dwt, fft and svd based watermarking technique for different
  wavelet transforms,'' \emph{Int. Journal of Scientific Research in Computer
  Science and Engineering}, p.~7, 2017.

\bibitem[Shehab and Batarfi(2017)]{stackcookies}
D.~A.-H. Shehab and O.~A. Batarfi, ``Rcr for preventing stack smashing attacks
  bypass stack canaries,'' in \emph{Computing Conference, 2017}.\hskip 1em plus
  0.5em minus 0.4em\relax IEEE, 2017, pp. 795--800.

\bibitem[Dang et~al.(2015)Dang, Maniatis, and
  Wagner]{stackshadow-and-cannariesp-perform}
T.~H. Dang, P.~Maniatis, and D.~Wagner, ``The performance cost of shadow stacks
  and stack canaries,'' in \emph{Proceedings of the 10th ACM Symposium on
  Information, Computer and Communications Security}.\hskip 1em plus 0.5em
  minus 0.4em\relax ACM, 2015, pp. 555--566.

\bibitem[Marco-Gisbert and Ripoll(2013)]{canary-brute-force}
H.~Marco-Gisbert and I.~Ripoll, ``Preventing brute force attacks against stack
  canary protection on networking servers,'' in \emph{Network Computing and
  Applications (NCA), 2013 12th IEEE International Symposium on}.\hskip 1em
  plus 0.5em minus 0.4em\relax IEEE, 2013, pp. 243--250.

\bibitem[Zhu et~al.(2017)Zhu, Zhou, Wang, Mu, and
  Mao]{byte-for-byte-stack-cooki}
J.~Zhu, W.~Zhou, Z.~Wang, D.~Mu, and B.~Mao, ``Diffguard: Obscuring sensitive
  information in canary based protections,'' in \emph{International Conference
  on Security and Privacy in Communication Systems}.\hskip 1em plus 0.5em minus
  0.4em\relax Springer, 2017, pp. 738--751.

\bibitem[Kilic et~al.(2014)Kilic, Kittel, and Eckert]{format-string}
F.~Kilic, T.~Kittel, and C.~Eckert, ``Blind format string attacks,'' in
  \emph{International Conference on Security and Privacy in Communication
  Systems}.\hskip 1em plus 0.5em minus 0.4em\relax Springer, 2014, pp.
  301--314.

\bibitem[Stojanovski et~al.(2007)Stojanovski, Gusev, Gligoroski, and
  Knapskog]{dep-virtual-alloc}
N.~Stojanovski, M.~Gusev, D.~Gligoroski, and S.~J. Knapskog, ``Bypassing data
  execution prevention on microsoftwindows xp sp2,'' in \emph{Availability,
  Reliability and Security, 2007. ARES 2007. The Second International
  Conference on}.\hskip 1em plus 0.5em minus 0.4em\relax IEEE, 2007, pp.
  1222--1226.

\bibitem[Onarlioglu et~al.(2010)Onarlioglu, Bilge, Lanzi, Balzarotti, and
  Kirda]{g-free}
K.~Onarlioglu, L.~Bilge, A.~Lanzi, D.~Balzarotti, and E.~Kirda, ``G-free:
  defeating return-oriented programming through gadget-less binaries,'' in
  \emph{Proceedings of the 26th Annual Computer Security Applications
  Conference}.\hskip 1em plus 0.5em minus 0.4em\relax ACM, 2010, pp. 49--58.

\bibitem[Li et~al.(2010)Li, Wang, Jiang, Grace, and Bahram]{retless}
J.~Li, Z.~Wang, X.~Jiang, M.~Grace, and S.~Bahram, ``Defeating return-oriented
  rootkits with return-less kernels,'' in \emph{Proceedings of the 5th European
  conference on Computer systems}.\hskip 1em plus 0.5em minus 0.4em\relax ACM,
  2010, pp. 195--208.

\bibitem[Davi et~al.(2011)Davi, Sadeghi, and Winandy]{ropdefender}
L.~Davi, A.-R. Sadeghi, and M.~Winandy, ``Ropdefender: A detection tool to
  defend against return-oriented programming attacks,'' in \emph{Proceedings of
  the 6th ACM Symposium on Information, Computer and Communications
  Security}.\hskip 1em plus 0.5em minus 0.4em\relax ACM, 2011, pp. 40--51.

\bibitem[Chen et~al.(2009)Chen, Xiao, Shen, Yin, Mao, and Xie]{drop}
P.~Chen, H.~Xiao, X.~Shen, X.~Yin, B.~Mao, and L.~Xie, ``Drop: Detecting
  return-oriented programming malicious code,'' in \emph{International
  Conference on Information Systems Security}.\hskip 1em plus 0.5em minus
  0.4em\relax Springer, 2009, pp. 163--177.

\bibitem[Jacobson et~al.(2014)Jacobson, Bernat, Williams, and Miller]{ropstop}
E.~R. Jacobson, A.~R. Bernat, W.~R. Williams, and B.~P. Miller, ``Detecting
  code reuse attacks with a model of conformant program execution,'' in
  \emph{International Symposium on Engineering Secure Software and
  Systems}.\hskip 1em plus 0.5em minus 0.4em\relax Springer, 2014, pp. 1--18.

\bibitem[Cheng et~al.(2014)Cheng, Zhou, Miao, Ding, Deng, et~al.]{ropecker}
Y.~Cheng, Z.~Zhou, Y.~Miao, X.~Ding, H.~Deng \emph{et~al.}, ``Ropecker: A
  generic and practical approach for defending against rop attack,'' 2014.

\bibitem[Pappas(2012)]{kbouncer}
V.~Pappas, ``kbouncer: Efficient and transparent rop mitigation,'' \emph{Apr},
  vol.~1, pp. 1--2, 2012.

\bibitem[Schuster et~al.(2014)Schuster, Tendyck, Pewny, Maa{\ss}, Steegmanns,
  Contag, and Holz]{antirop}
F.~Schuster, T.~Tendyck, J.~Pewny, A.~Maa{\ss}, M.~Steegmanns, M.~Contag, and
  T.~Holz, ``Evaluating the effectiveness of current anti-rop defenses,'' in
  \emph{International Workshop on Recent Advances in Intrusion
  Detection}.\hskip 1em plus 0.5em minus 0.4em\relax Springer, 2014, pp.
  88--108.

\bibitem[Bletsch et~al.(2011{\natexlab{b}})Bletsch, Jiang, and Freeh]{cfl}
T.~Bletsch, X.~Jiang, and V.~Freeh, ``Mitigating code-reuse attacks with
  control-flow locking,'' in \emph{Proceedings of the 27th Annual Computer
  Security Applications Conference}.\hskip 1em plus 0.5em minus 0.4em\relax
  ACM, 2011, pp. 353--362.

\bibitem[Hiser et~al.(2012)Hiser, Nguyen-Tuong, Co, Hall, and Davidson]{ilr}
J.~Hiser, A.~Nguyen-Tuong, M.~Co, M.~Hall, and J.~W. Davidson, ``Ilr: Where'd
  my gadgets go?'' in \emph{Security and Privacy (SP), 2012 IEEE Symposium
  on}.\hskip 1em plus 0.5em minus 0.4em\relax IEEE, 2012, pp. 571--585.

\bibitem[Triantafyllis et~al.(2003)Triantafyllis, Vachharajani, Vachharajani,
  and August]{compiler-optim}
S.~Triantafyllis, M.~Vachharajani, N.~Vachharajani, and D.~I. August,
  ``Compiler optimization-space exploration,'' in \emph{Code Generation and
  Optimization, 2003. CGO 2003. International Symposium on}.\hskip 1em plus
  0.5em minus 0.4em\relax IEEE, 2003, pp. 204--215.

\bibitem[Microsoft(2015)]{naked}
\BIBentryALTinterwordspacing
Microsoft. (2015) Microsoft msdn, c++ naked functions. [Online]. Available:
  \url{https://msdn.microsoft.com/en-us/library/h5w10wxs.aspx}
\BIBentrySTDinterwordspacing

\bibitem[Denning(2006)]{locality}
P.~J. Denning, ``The locality principle,'' in \emph{Communication Networks And
  Computer Systems: A Tribute to Professor Erol Gelenbe}.\hskip 1em plus 0.5em
  minus 0.4em\relax World Scientific, 2006, pp. 43--67.

\bibitem[Kleen et~al.(2005)Kleen, Stienberg, Anschel, Sibony, and
  Greenberg]{locality-poor}
A.~Kleen, E.~Stienberg, M.~Anschel, Y.~Sibony, and S.~Greenberg, ``An improved
  instruction cache replacement algorithm,'' in \emph{Signal Processing Systems
  Design and Implementation, 2005. IEEE Workshop on}.\hskip 1em plus 0.5em
  minus 0.4em\relax IEEE, 2005, pp. 573--578.

\bibitem[Agner(2008)]{optimizing}
F.~Agner, ``Optimizing subroutines in assembly language: An optimization guide
  for x86 platforms,'' 2008.

\bibitem[Microsoft(2005)]{windbg}
\BIBentryALTinterwordspacing
Microsoft. (2005) Debugging tools for windows (windbg, kd, cdb, ntsd).
  [Online]. Available:
  \url{https://docs.microsoft.com/en-us/windows-hardware/drivers/debugger/}
\BIBentrySTDinterwordspacing

\bibitem[Rajwar et~al.(2013)Rajwar, Lachner, Knauth, and Lai]{LBR}
R.~Rajwar, P.~Lachner, L.~A. Knauth, and K.~K. Lai, ``Processor with last
  branch record register storing transaction indicator,'' Jul.~2 2013, uS
  Patent 8,479,053.

\bibitem[Naftaly(2012)]{pin}
S.~Naftaly, ``Pin-a dynamic binary instrumentation tool,'' 2012.

\bibitem[Salwan(2011)]{ropgadget}
J.~Salwan, ``Ropgadget--gadgets finder and auto-roper,'' 2011.

\bibitem[Bhattacharya et~al.(2016)Bhattacharya, Men{\'e}ndez, Barr, and
  Clark]{accurancy-best}
S.~Bhattacharya, H.~D. Men{\'e}ndez, E.~Barr, and D.~Clark, ``Itect: Scalable
  information theoretic similarity for malware detection,'' \emph{arXiv
  preprint arXiv:1609.02404}, 2016.

\end{thebibliography}

\end{document}